\documentclass[11pt,a4paper]{article}
\usepackage{amsmath,amsthm,amsfonts,amssymb,mathrsfs}
\usepackage{graphicx}
\usepackage{subfigure}
\usepackage[latin1]{inputenc}
\usepackage[swedish,english]{babel}
\renewcommand{\vec}[1]{
{\boldsymbol#1}}

\begin{document}
\title{Determination of normalized magnetic eigenfields in microwave
    cavities}
\author{Johan Helsing\thanks{Centre for Mathematical Sciences, Lund
    University, Sweden}~~and Anders Karlsson\thanks{Electrical and
    Information Technology, Lund University, Sweden}}
\date{\today}
\maketitle

\begin{abstract}
  The magnetic field integral equation for axially symmetric cavities
  with perfectly conducting surfaces is discretized according to a
  high-order convergent Fourier--Nyström scheme. The resulting solver
  is used to determine eigenwavenumbers and normalized magnetic
  eigenfields to very high accuracy in the entire computational
  domain.
\end{abstract}

\section{Introduction}

This work is on the numerical solution of the time harmonic Maxwell
equations in axially symmetric hollow microwave cavities with smooth
and perfectly electric conducting (PEC) surfaces. We use the magnetic
field integral equation (MFIE) and high-order convergent
Fourier--Nyström discretization to find normalized magnetic
eigenfields to high accuracy.

Particle accelerators are the most common application for microwave
cavities, also known as Radio Frequency (RF) cavities. Today, all high
energy particle accelerators use such cavities for particle
acceleration by means of eigenfields, excited by external sources. The
cavities consist of one, or several, axially symmetric cells, where in
each cell the excited eigenfield corresponds to the TM$_{010}$ mode in
a cylindrical cavity, see~\cite[Chapter 1]{Wangler08}. In addition to
the eigenfields excited by the external sources, there are wakefields
that consist of a large number of higher-order modes (HOM). The
wakefields are excited in cavities and flanges by the beam of
particles. See~\cite[Chapter 11]{Wangler08} for a more detailed
discussion. The wakefields affect the trajectories of the particles
and, by that, the quality of the beam. To prevent that harmful
wakefields are excited in the accelerator, numerical simulations are
used both in its design process and during operation. These
simulations need to cover a frequency band up to at least 10 times the
frequency of the external source. At the highest frequency the width
of each cell of a cavity is on the order of 5-10 wavelengths and the
length on the order of 3-6 wavelengths. Accelerators can have cavities
with as many as 20 cells. In the collaboration of the present authors
with scientists at the synchrotron light source MAX IV and the
European Spallation Source (ESS), both under construction in Lund,
Sweden, we see a great need for improved numerical tools for accurate
wakefield evaluation at higher frequencies. This is a motivation for
our work.

During the last 30 years, software packages have become increasingly
important for the design of microwave cavities. Today, finite element
method (FEM) based software packages like COMSOL Multiphysics, ANSYS,
and HFSS, finite difference method (FD and FDTD) based packages like
SUPERFISH \cite{HalbachHolsinger76} and GdfidL \cite{Bruns98}, and
finite integration technique (FIT) \cite{Weiland77} based packages
like MAFIA and CST, are common design tools. All of these packages
rely on a partial differential equation (PDE) formulation of the time
harmonic Maxwell equations and they discretize the volume, or, in the
axially symmetric case, the cross section of the cavity. They are
suitable for evaluations of low-order modes in cavities, but less
suitable for accurate evaluations of high-frequency wakefields, due to
their relatively low-order convergence.

To our knowledge, there are no published papers on the MFIE applied to
microwave cavities. Neither have we found any published benchmark
results. Authors using the MFIE in axially symmetric domains apply it
to exterior problems and the same is true for the related electric
field integral equation (EFIE) and the combined field integral
equation (CFIE). The method of moments (MoM) is the most common method
for discretization, see \cite{Andreasen1965}, \cite{Gedney+Mittra90},
\cite{Mautz+Harrington69}, and the reference list in
\cite{Tong+Chew08}. Only a few papers favor Nyström methods, see
\cite{Flemingetal05} and \cite{Vicoetal13}. The CFIE is often used for
exterior problems since, in contrast to MFIE and EFIE, it provides
unique solutions also at the eigenwavenumbers of the interior problem.
For the interior problem, CFIE is unnecessarily complicated. An
alternative to CFIE is presented in \cite{Epstein10} and
\cite{Epstein13}, where two coupled surface integral equations with
unique solutions are derived by introducing surface potentials,
referred to as Debye sources.

The present work shows that high-order convergent Nyström schemes for
the MFIE can be efficiently implemented when applied to cavities with
PEC surfaces. Very accurate results are obtained both for surface
current densities and for surface charge densities at a broad range of
eigenwavenumbers. These results can, via a post-processor, be carried
over to normalized magnetic eigenfields at all positions inside the
cavity, also close to surfaces where integral equation techniques
usually encounter difficulties. Our post-processor takes advantage of
a new surface integral expression for the normalization. In terms of
discretization techniques used, we rely solely on~\cite{HelsKarl14}.
In~\cite{HelsKarl14} we developed an explicit kernel-split panel-based
Fourier--Nyström scheme for integral equations on axially symmetric
surfaces where the integral operators can have weakly singular or
Cauchy-type singular kernels. The numerical examples
in~\cite{HelsKarl14} deal with acoustic eigenfields and only require
the discretization of three distinct integral operators. The numerical
examples of the present work involve around 30 distinct, but very
similar, integral operators. They are all discretized using techniques
from~\cite{HelsKarl14}. See~\cite{Youn12} for a related Nyström
scheme, without kernel splits, in the acoustic setting.

The paper is organized as follows: Section~\ref{sec:prob} presents the
MFIE and an integral representation of the magnetic field in a concise
notation. Section~\ref{sec:Fourier} defines an azimuthal Fourier
transformation of $2\pi$-periodic functions and applies it to the MFIE
and to the field representation. Section~\ref{sec:discrete} reviews
our Nyström discretization scheme for transformed integral operators
and equations. Section~\ref{sec:numerical} contains numerical examples
with relevance to accelerator technology and nano-optics. The
conclusions in Section~\ref{sec:conclus} relate to future research
directions. In order to maintain a high narrative pace in the main
body of the paper we have collected some rather important details in
three appendices. Appendix~\ref{app:A} contains a derivation of the
surface integral expression used for the normalization.
Appendix~\ref{app:B} presents a number of useful relations between
various layer densities and potentials. Appendix~\ref{app:toro} gives
a short {\sc Matlab} code for the accurate evaluation of two
half-integer Legendre functions.

\section{Problem formulation} 
\label{sec:prob}

This section introduces the MFIE for the time harmonic Maxwell
equations in a notation that is particularly adapted to axially
symmetric hollow cavities with PEC surfaces. Most of the material is
well known, see
\cite{Flemingetal05,Gedney+Mittra90,GlissonWilton79,Mautz+Harrington69}.

\subsection{Basic notation}

\begin{figure}
 \centering 
 \includegraphics[height=47mm]{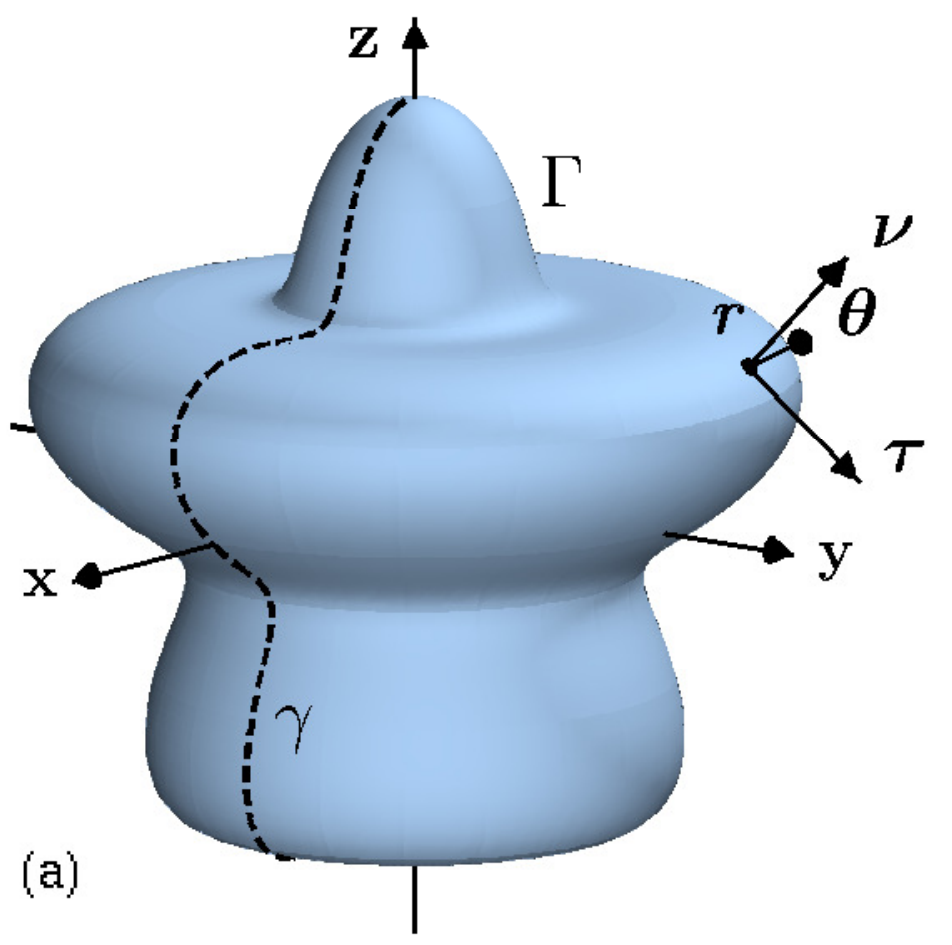}
 \includegraphics[height=47mm]{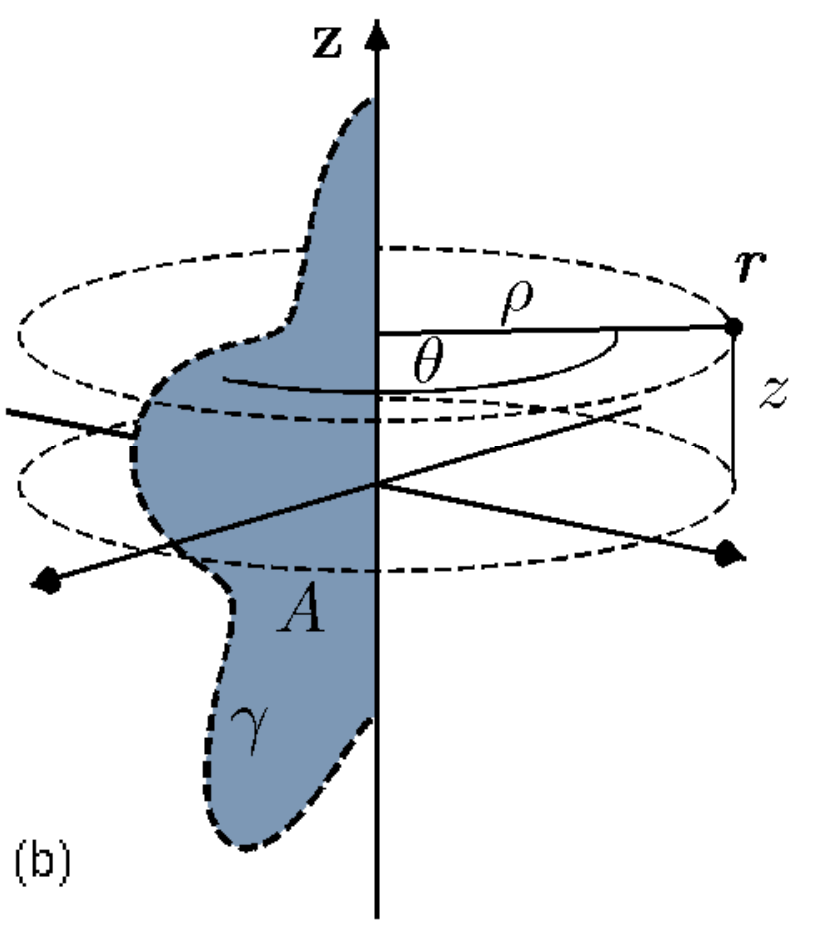}
 \includegraphics[height=47mm]{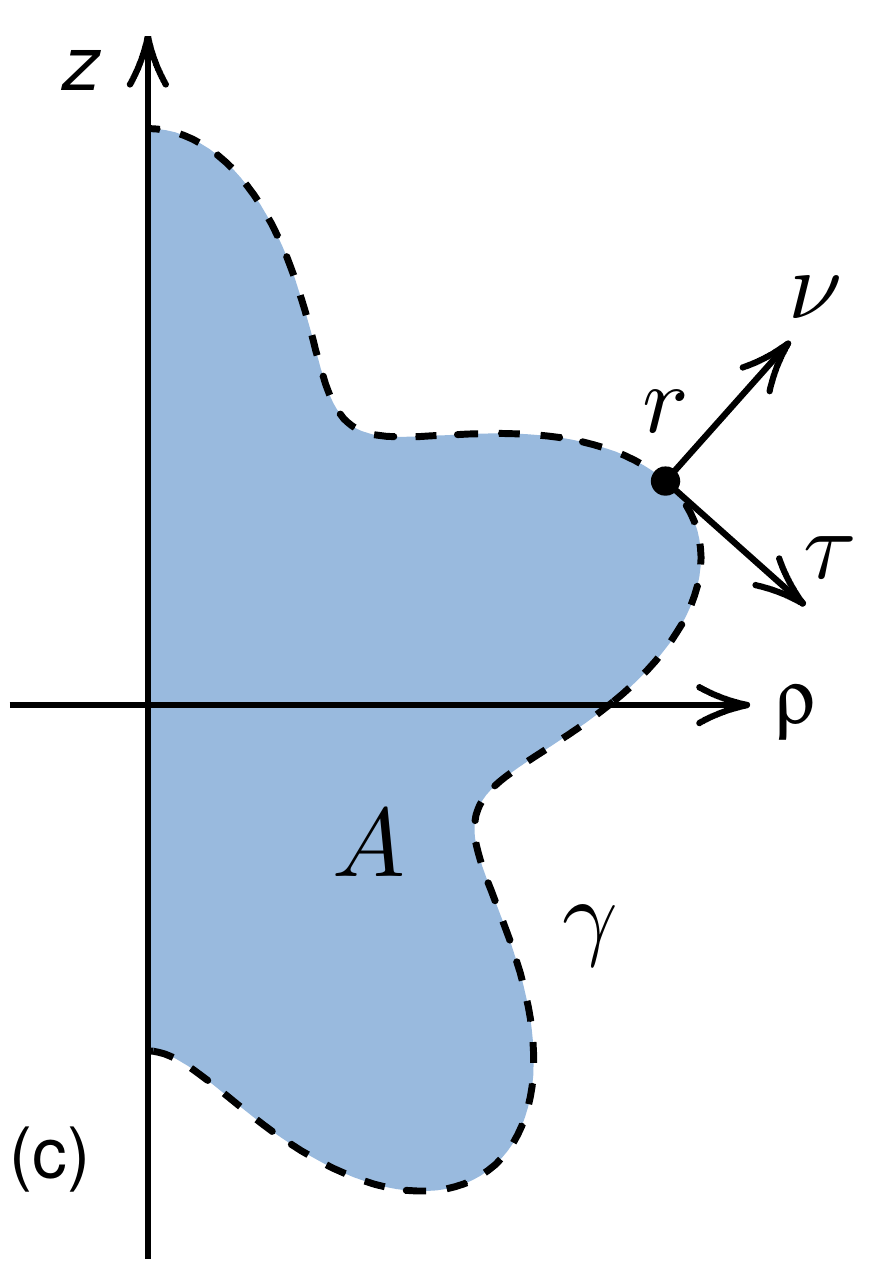}
\caption{\sf An axisymmetric surface $\Gamma$ generated by a curve 
  $\gamma$. (a) A point ${\vec r}$ on $\Gamma$ has outward unit normal
  $\boldsymbol{\nu}$ and tangent vector $\boldsymbol{\tau}$. (b)
  ${\vec r}$ has radial distance $\rho$, azimuthal angle $\theta$, and
  height $z$. The planar domain $A$ is bounded by $\gamma$ and the
  $z$-axis. (c) Coordinate axes and vectors in the half-plane
  $\mathbb{R}^{2+}$.}
\label{fig:geometry}
\end{figure}

Let $\Gamma$ be an axially symmetric surface enclosing a
three-dimensional domain $V$ (a body of revolution) and let
\begin{equation}
\vec r=(x,y,z)=(\rho\cos\theta,\rho\sin\theta,z)
\end{equation}
denote a point in $\mathbb{R}^3$. Here $\rho=\sqrt{x^2+y^2}$ is the
distance from the $z$-axis and $\theta$ is the azimuthal angle. The
outward unit normal $\vec\nu$ at a point $\vec r$ on $\Gamma$ is defined as
\begin{equation}
\vec\nu=(\nu_{\rho }\cos\theta,\nu_{\rho }\sin\theta,\nu_z)\,.\\
\end{equation}
We also need the unit vectors
\begin{align}
\vec\rho&=(\cos\theta,\sin\theta,0)\,,\\
\vec\theta&=(-\sin\theta,\cos\theta,0)\,,\\
\vec\tau&=\vec \theta\times\vec \nu
 =(\nu_z\cos\theta, \nu_z\sin\theta,-\nu_{\rho })\,,\\
\vec z&=(0,0,1)\,,
\end{align}
where $\vec\theta$ and $\vec\tau$ are two tangential unit vectors.
See Figure~\ref{fig:geometry}(a) and~\ref{fig:geometry}(b).

The angle $\theta=0$ defines a half-plane $\mathbb{R}^{2+}$ in
$\mathbb{R}^3$ whose intersection with $\Gamma$ corresponds to a
generating curve $\gamma$. Let $r=(\rho,z)$ be a point in
$\mathbb{R}^{2+}$ and let $A$ be the planar domain bounded by $\gamma$
and the $z$-axis. The outward unit normal on $\gamma$ is
$\nu=(\nu_\rho,\nu_z)$ and $\tau=(\nu_z,-\nu_\rho)$ is a tangent. See
Figure~\ref{fig:geometry}(b) and~\ref{fig:geometry}(c). The unit
vectors in the $\rho$- and $z$-directions are $\hat{\rho}$ and
$\hat{z}$.

\subsection{PDE formulation}

Our primary interest is the magnetic field. With vacuum in $V$ and
with $\Gamma$ perfectly conducting, the magnetic field $\vec H(\vec
r)$ satisfies the following system of partial differential equations
\begin{align}
\nabla^2\vec H(\vec r)+k^2\vec H(\vec r)&=\vec 0\,,\qquad\vec r\in V\,,
\label{PDE}\\
\nabla\cdot\vec H(\vec r)&=0\,,\qquad\vec r\in V\,,
\label{eq:XTRA}
\end{align}
with boundary condition
\begin{equation}
\lim_{V\ni \vec r\to \vec r^\circ}
\vec\nu^\circ\times(\nabla\times\vec H(\vec r))=\vec 0\,, 
\qquad\vec r^\circ\in\Gamma\,.
\label{eq:BC}
\end{equation}
We will find nontrivial solutions to these equations in a fast and
accurate fashion via the MFIE.

The values $k^2$ for which the system~(\ref{PDE}), (\ref{eq:XTRA}),
and~(\ref{eq:BC}) admits nontrivial solutions are called eigenvalues.
We refer to the corresponding fields $\vec H(\vec r)$ as magnetic
eigenfields and to $k$ as eigenwavenumbers. The eigenvalues constitute
a real, positive and countable set, accumulating only at
infinity~\cite{Wen08}. The eigenvalues have finite multiplicity.
Magnetic eigenfields that correspond to distinct eigenvalues are
orthogonal and subspaces of magnetic eigenfields that correspond to
particular eigenvalues can be given orthogonal bases.

We introduce an inner product on $V$, denoted by angle brackets
$\langle\cdot,\cdot\rangle$, and an induced norm
\begin{equation}
\langle \vec F, \vec G\rangle=
\int_V \vec F^\ast(\vec r)\cdot\vec G(\vec r)\,{\rm d}V\,,
\qquad
\|\vec F\|^2=\langle \vec F, \vec F\rangle\,.
\label{eq:Hnormdef}
\end{equation}
Here $\vec F(\vec r)$ and $\vec G(\vec r)$ are vector (or scalar)
fields on $V$ and the symbol '$\ast$' denotes the complex conjugate.
Magnetic eigenfields $\vec H(\vec r)$ are normalized so that
\begin{equation}
\|\vec H\|^2\equiv
\int_V\vec H^\ast(\vec r)\cdot\vec H(\vec r)\,{\rm d}V=1\,.
\label{eq:Hnorm}
\end{equation}
The volume integral in \eqref{eq:Hnorm} is referred to as {\em the
  normalization integral}. In Appendix~\ref{app:A} it is reformulated
as a surface integral that is well suited for numerical evaluation in
the framework of the MFIE.

\subsection{The MFIE}

The magnetic field is represented in terms of the surface current
density
\begin{equation}
\vec J_{\rm s}(\vec r^\circ)=
\lim_{V\ni \vec r \to \vec r^\circ}-\vec\nu^\circ\times\vec H(\vec r)\,,
\qquad \vec r^\circ\in\Gamma\,,
\label{eq:JSdef}
\end{equation}
by the Stratton--Chu integral
representation~\cite{StrattonChu1939,Waterman1971}. When~(\ref{eq:BC})
holds, this representation assumes the form
\begin{equation}
\left.
\begin{aligned}
&\vec H(\vec r)\\
&\vec 0
\end{aligned}
\right\}=
\nabla\times\int_\Gamma\vec J_{\rm s}(\vec r')
\Phi_k(\vec r,\vec r')
\,{\rm d}\Gamma'\,\,\left\{
\begin{aligned}
&\vec r\in V\,,\\
&\vec r\in\mathbb{R}^3\setminus V\cup\Gamma\,,
\end{aligned}
\right.
\label{eq:Hrep}
\end{equation}
where, using the time dependence $e^{-{\rm i}\omega t}$,
\begin{equation}
\Phi_k(\vec r,\vec r')=\frac{e^{{\rm i}k\vert\vec r-\vec r'\vert}}
{4\pi\vert\vec r-\vec r'\vert}
\end{equation}
is the causal fundamental solution to the Helmholtz equation. The
upper equation in \eqref{eq:Hrep} gives the magnetic field in $V$. The
lower equation states that $\vec J_{\rm s}(\vec r)$ induces a zero
magnetic field outside $\Gamma$.
 
Taking the limit $V\ni\vec r\to\vec r^\circ\in\Gamma$ in
\eqref{eq:Hrep} and using (\ref{eq:JSdef}), one gets the MFIE:
\begin{equation}
\frac{1}{2}\vec J_{\rm s}(\vec
r)-\vec{\nu}\times\int_\Gamma\left(\vec J_{\rm s}(\vec
  r')\times\nabla
\Phi_k(\vec r,\vec r')\right)\,{\rm d}\Gamma'=\vec 0\,,
\qquad \vec r\in \Gamma\,.
\label{eq:MFIE}
\end{equation}
A solution to the MFIE, together with~(\ref{eq:Hrep}), is also a
solution to the system~(\ref{PDE}), (\ref{eq:XTRA}),
and~(\ref{eq:BC}). The kernel of the integral operator in the MFIE is
weakly singular, whereas the kernel in~(\ref{eq:Hrep}) exhibits a
stronger singularity in the limit $\mathbb{R}^3\setminus\Gamma\ni\vec
r\to\vec r^\circ\in\Gamma$.

In order to cast~(\ref{eq:MFIE}) in a form more suitable for numerical
solution we decompose $\vec J_{\rm s}(\vec r)$ in its tangential
components
\begin{equation}\label{eq:decompJ}
\vec J_{\rm s}(\vec r)=\vec\tau J_{\tau}(\vec r)+\vec\theta J_{\theta}(\vec r)
\end{equation}
and rewrite and split~(\ref{eq:MFIE}) into the two coupled scalar
equations
\begin{align}
\begin{split}
  J_{\tau}(\vec r)&-2\int_\Gamma\left((\vec\tau\cdot\vec\tau')
  J_{\tau}(\vec r')+(\vec\tau\cdot\vec\theta')J_{\theta}(\vec r')\right)
  \left(\vec\nu\cdot\nabla \Phi_k(\vec r,\vec r')\right)\,{\rm d}\Gamma'\\
 &                 +2\int_\Gamma\left((\vec\nu\cdot\vec\tau')
  J_{\tau}(\vec r')+(\vec\nu\cdot\vec\theta')J_{\theta}(\vec r')\right)
 \left(\vec\tau\cdot\nabla\Phi_k(\vec r,\vec r')\right)\,{\rm d}\Gamma'=0\,,\\
 J_{\theta}(\vec r)&-2\int_\Gamma\left((\vec\theta\cdot\vec\tau') 
   J_{\tau}(\vec r')+(\vec\theta\cdot\vec\theta')J_{\theta}(\vec r')\right)
  \left(\vec\nu\cdot\nabla\Phi_k(\vec r,\vec r')\right)\,{\rm d}\Gamma'\\
 &                  +2\int_\Gamma\left((\vec\nu \cdot\vec\tau') 
   J_{\tau}(\vec r')+(\vec\nu\cdot\vec\theta')J_{\theta}(\vec r')\right)
 \left(\vec\theta\cdot\nabla\Phi_k(\vec r,\vec r')\right)\,{\rm d}\Gamma'=0\,.
\end{split}
\label{eq:sys12}
\end{align}

It is convenient to express the system~(\ref{eq:sys12}) in the more
compact form
\begin{equation}
\begin{split}
\left(I+K_1\right)J_{\tau}(\vec r)+{\rm i}K_2J_{\theta}(\vec r)=0\,,
\qquad {\vec r}\in\Gamma\,, 
\\
{\rm i}K_3J_{\tau}(\vec r)+\left(I+K_4\right)J_{\theta}(\vec r)=0\,,
\qquad {\vec r}\in\Gamma\,.
\end{split}
\label{eq:comp12}
\end{equation}
Here $I$ is the identity. The double-layer type operators $K_\alpha$,
with kernels $K_\alpha(\vec r,\vec r')$, are defined by their actions
on a layer density $g({\vec r})$ on $\Gamma$ as
\begin{equation}
K_\alpha g({\vec r})
=\int_\Gamma K_\alpha(\vec r,\vec r')g({\vec r'})\,{\rm d}\Gamma'
=\int_\Gamma D_\alpha(\vec r,\vec r')
P(\vec r,\vec r')g({\vec r'})\,{\rm d}\Gamma'\,, 
\label{eq:Kdef}
\end{equation}
where $\alpha=1,2,3,4$,
\begin{align}
D_1(\vec r,\vec r')&= 2\frac{\left(\rho\nu_{\rho}'-(\nu'\cdot
    r'-\nu_z'z)\cos(\theta-\theta')\right)}{4\pi\vert\vec r-\vec r'\vert^3}\,, 
\label{D1}\\
D_2(\vec r,\vec r')&=-2{\rm i}
    \frac{(z-z')\sin(\theta-\theta')}{4\pi\vert\vec r-\vec r'\vert^3}\,,
\label{D2}\\
D_3(\vec r,\vec r')&=2{\rm i}\frac{(\nu_z'\nu\cdot r-\nu_z\nu'\cdot
  r')\sin(\theta-\theta')}{4\pi\vert\vec r-\vec r'\vert^3}\,,
\label{D3}\\
D_4(\vec r,\vec r')&=-2\frac{\left(\rho'\nu_{\rho}-(\nu\cdot
    r-\nu_zz')\cos(\theta-\theta')\right)}{4\pi\vert\vec r-\vec r'\vert^3}\,,
\label{D4}
\end{align}
and
\begin{align}
P(\vec r,\vec r')&=
(1-{\rm i}k\vert\vec r-\vec r'\vert)e^{{\rm i}k\vert\vec r-\vec r'\vert}\,,\\
\vert\vec r-\vec r'\vert&=
\sqrt{\rho^2+\rho'^2-2\rho\rho'\cos(\theta-\theta')+(z-z')^2}\,.
\end{align}
The functions $D_\alpha(\vec r,\vec r')$ can be viewed as static
kernels, corresponding to wavenumber $k=0$.

\subsection{Expressions for the magnetic field}
\label{sec:magexp}

The magnetic field~(\ref{eq:Hrep}) can be expressed in a form
analogous to~(\ref{eq:comp12}), which is better suited for numerics.
For this, we introduce the decomposition
\begin{equation}
\vec H(\vec r)=\vec\rho H_{\rho}(\vec r)
+\vec\theta H_{\theta}(\vec r)+\vec zH_z(\vec r)\,.
\end{equation}
Straightforward calculations give for $\vec r\in\mathbb{R}^3\setminus\Gamma$
\begin{equation}
\begin{split}
H_{\rho}(\vec r)&={\rm i}K_5J_{\tau}(\vec r)+K_6J_{\theta}(\vec r)\,,\\
H_{\theta}(\vec r)&=K_7J_{\tau}(\vec r)+{\rm i}K_8J_{\theta}(\vec r)\,,\\
H_{z}(\vec r)&={\rm i}K_9J_{\tau}(\vec r)+K_{10}J_{\theta}(\vec r)\,,
\end{split}
\label{eq:comp13}
\end{equation}
where $K_\alpha$, $\alpha=5,6,7,8,9,10$, are defined as
in~(\ref{eq:Kdef}) with
\begin{align}
D_5(\vec r,\vec r')&=-{\rm i}\frac{\left(\nu'\cdot
    r'-\nu_z'z\right)\sin(\theta-\theta')}{4\pi\vert\vec r-\vec r'\vert^3}\,,
\label{D5}\\
D_6(\vec r,\vec r')&=\frac{(z-z')\cos(\theta-\theta')}
{4\pi\vert\vec r-\vec r'\vert^3}\,,
\label{D6}\\
D_7(\vec r,\vec r')&=-\frac{\left(\nu_{\rho}'\rho-\left(\nu'\cdot
    r'-\nu_z'z\right)
\cos(\theta-\theta')\right)}{4\pi\vert\vec r-\vec r'\vert^3}\,,
\label{D7}\\
D_8(\vec r,\vec r')&={\rm i}
\frac{(z-z')\sin(\theta-\theta')}{4\pi\vert\vec r-\vec r'\vert^3}\,,
\label{D8}\\
D_9(\vec r,\vec r')&=-{\rm i}
\frac{\nu_z'\rho\sin(\theta-\theta')}{4\pi\vert\vec r-\vec r'\vert^3}\,,
\label{D9}\\
D_{10}(\vec r,\vec r')&=
\frac{\rho'-\rho\cos(\theta-\theta')}
{4\pi\vert\vec r-\vec r'\vert^3}\,.
\label{D10}
\end{align}

\section{Fourier series expansions}
\label{sec:Fourier}

The aim of this paper is to present a high-order convergent and
accurate discretization scheme to solve the MFIE and to to evaluate
magnetic eigenfields, normalized by~(\ref{eq:Hnorm}). We employ a
Fourier--Nyström technique where the first step is an azimuthal
Fourier transformation of the MFIE system~(\ref{eq:comp12}) and of the
system for the decomposed magnetic field~(\ref{eq:comp13}).

Several 2$\pi$-periodic quantities need to be expanded. We define the
azimuthal Fourier coefficients 
\begin{align}
g_n(r)&=\frac{1}{\sqrt{2\pi}}\int_{-\pi}^{\pi}e^{-{\rm i}n\theta}
g({\vec r})\,{\rm d}\theta\,, 
\label{eq:gF}\\
G_n(r,r')&=
\frac{1}{\sqrt{2\pi}}\int_{-\pi}^{\pi}e^{-{\rm i}n(\theta-\theta')}
G({\vec r},{\vec r}')\,{\rm d}(\theta-\theta')\,,
\label{eq:GF}
\end{align}
where $g(\vec r)$ can represent functions like $J_\tau(\vec r)$,
$J_\theta(\vec r)$, $H_\rho(\vec r)$, $H_\theta(\vec r)$, and
$H_z(\vec r)$ and where $G({\vec r},{\vec r}')$ can represent
functions like $K_\alpha({\vec r},{\vec r}')$, $D_\alpha({\vec
  r},{\vec r}')$, and $P({\vec r},{\vec r}')$. The subscript $n$ is
the azimuthal index. The coefficients $G_n(r,r')$ may also be called
transformed kernels or modal Green's functions.

Expansion and integration of~(\ref{eq:comp12}) and~(\ref{eq:comp13})
over $\theta'$ give the system of modal integral equations
\begin{equation}
\begin{split}
\left(I+\sqrt{2\pi}K_{1n}\right)J_{\tau n}(r)
+{\rm i}\sqrt{2\pi}K_{2n}J_{\theta n}(r)=0\,, \qquad r\in\gamma\,, \\
 {\rm i}\sqrt{2\pi}K_{3n}J_{\tau n}(r)+
\left(I+\sqrt{2\pi}K_{4n}\right)J_{\theta n}(r)=0\,, \qquad r\in\gamma\,,
\end{split}
\label{eq:comp12F}
\end{equation}
and the modal representation of the magnetic field for
$r\in\mathbb{R}^{2+}\setminus\gamma$
\begin{equation}
\begin{split}
H_{\rho n}(r)  &={\rm i}\sqrt{2\pi}K_{5n}J_{\tau n}(r)
                     +\sqrt{2\pi}K_{6n}J_{\theta n}(r)\,,\\
H_{\theta n}(r)&=\sqrt{2\pi}K_{7n}J_{\tau n}(r)
                     +{\rm i}\sqrt{2\pi}K_{8n}J_{\theta n}(r)\,,\\
H_{z n}(r)     &={\rm i}\sqrt{2\pi}K_{9n}J_{\tau n}(r)
                     +\sqrt{2\pi}K_{10n}J_{\theta n}(r)\,.
\end{split}
\label{eq:comp13F}
\end{equation}
The azimuthal index is
$n=0,\pm 1,\pm 2,\ldots$ and
\begin{equation}
K_{\alpha n}g_n(r)=
\int_\gamma K_{\alpha n}(r,r')g_n(r')\rho'\,{\rm d}\gamma'\,, 
\qquad \alpha=1,\ldots,10\,.
\label{eq:Kndef}
\end{equation}

\subsection{Azimuthal Fourier coefficients in closed form}
\label{sec:azimuthn}

When $r$ and $r'$ are far apart, the kernels $K_\alpha(\vec r,\vec
r')$ are smooth functions and the $K_{\alpha n}(r,r')$, present
in~(\ref{eq:comp12F}) and~(\ref{eq:comp13F}), can be efficiently
evaluated from~(\ref{eq:GF}) using discrete Fourier transform
techniques (FFT). When $r\approx r'$, this is not true. Then it is
more economical to split each $K_\alpha(\vec r,\vec r')$ into two
terms: a smooth term, which is transformed via FFT, and a non-smooth
term, which is transformed by convolution of $D_{\alpha n}(r,r')$ with
parts of $P_n(r,r')$. See~\cite[Section 6]{HelsKarl14} for details.
We use $2N+1-n$ terms in the convolutions, where $N$ is an integer
controlling all FFT operations.

The coefficients $D_{\alpha n}(r,r')$, for $r\approx r'$, are also costly to
evaluate from~(\ref{eq:GF}). Fortunately, the $D_{\alpha n}(r,r')$ can be
obtained as closed-form expressions involving half-integer degree
Legendre functions of the second kind
\begin{equation}
\mathfrak{Q}_{n-\frac{1}{2}}(\chi)=
\int_{-\pi}^{\pi}\frac{\cos(nt)\,{\rm d}t}
{\sqrt{8\left(\chi-\cos(t)\right)}}\,,
\label{eq:toro}
\end{equation}
which are cheap to evaluate. The functions
$\mathfrak{Q}_{n-\frac{1}{2}}(\chi)$, with real arguments $\chi\ge 1$,
may also be called toroidal harmonics~\cite{Segu00}. They are
symmetric with respect to $n$ and exhibit logarithmic singularities at
$\chi=1$. Introducing
\begin{gather}
\eta=\left(8\pi^3\rho\rho'\right)^{-\frac{1}{2}}\,,
\label{eq:etadef}\\
\chi=1+\frac{|r-r'|^2}{2\rho\rho'}\,,
\label{eq:chidef}\\
d(\nu)=\frac{\nu\cdot(r-r')}{|r-r'|^2}\,,
\label{eq:ddef}\\
\mathfrak{R}_n(\chi)=\frac{2n-1}{\chi+1}\left(
\chi\mathfrak{Q}_{n-\frac{1}{2}}(\chi)
-\mathfrak{Q}_{n-\frac{3}{2}}(\chi)\right)\,,
\label{eq:Rfrak}
\end{gather}
one can write
\begin{align}
D_{1n}(r,r')&=-2\eta\left[
d(\nu')\mathfrak{R}_n(\chi)-\frac{\left(\nu'\cdot r'-\nu'_zz\right)}
{2\rho\rho'}
\left(\mathfrak{R}_n(\chi)+\mathfrak{Q}_{n-\frac{1}{2}}(\chi)\right)\right]\,,
\label{eq:D1n}\\
D_{2n}(r,r')&=-2\eta
\frac{\left(z-z'\right)}{\rho\rho'}
n\mathfrak{Q}_{n-\frac{1}{2}}(\chi)\,,
\label{eq:D2n}\\
D_{3n}(r,r')&=2\eta
\frac{\left(\nu_z'\nu\cdot r-\nu_z\nu'\cdot r'\right)}{\rho\rho'}
n\mathfrak{Q}_{n-\frac{1}{2}}(\chi)\,,
\label{eq:D3n}\\
D_{4n}(r,r')&=-2\eta\left[
d(\nu)\mathfrak{R}_n(\chi)+\frac{\left(\nu\cdot r-\nu_zz'\right)}{2\rho\rho'}
\left(\mathfrak{R}_n(\chi)+\mathfrak{Q}_{n-\frac{1}{2}}(\chi)\right)\right]\,,
\label{eq:D4n}
\end{align}
and
\begin{align}
D_{5n}(r,r')&=-\eta\frac{\left(\nu'\cdot r'-\nu_z'z\right)}{\rho\rho'}
n\mathfrak{Q}_{n-\frac{1}{2}}(\chi)\,,
\label{eq:D5n}\\
D_{6n}(r,r')&=-\eta\left[
d(\hat{z})\mathfrak{R}_n(\chi)+\frac{\left(z-z'\right)}{2\rho\rho'}
\left(\mathfrak{R}_n(\chi)+\mathfrak{Q}_{n-\frac{1}{2}}(\chi)\right)\right]\,,
\label{eq:D6n}\\
D_{7n}(r,r')&=\eta\left[
d(\nu')\mathfrak{R}_n(\chi)-\frac{\left(\nu'\cdot r'-\nu'_zz\right)}
{2\rho\rho'}
\left(\mathfrak{R}_n(\chi)+\mathfrak{Q}_{n-\frac{1}{2}}(\chi)\right)\right]\,,
\label{eq:D7n}\\
D_{8n}(r,r')&=\eta
\frac{\left(z-z'\right)}{\rho\rho'}n\mathfrak{Q}_{n-\frac{1}{2}}(\chi)\,,
\label{eq:D8n}\\
D_{9n}(r,r')&=-\eta\frac{\nu_z'}{\rho'}n\mathfrak{Q}_{n-\frac{1}{2}}(\chi)\,,
\label{eq:D9n}\\
D_{10n}(r,r')&=\eta\left[
d(\hat{\rho})\mathfrak{R}_n(\chi)+\frac{1}{2\rho'}
\left(\mathfrak{R}_n(\chi)+\mathfrak{Q}_{n-\frac{1}{2}}(\chi)\right)\right]\,.
\label{eq:D10n}
\end{align}

Our derivation of~(\ref{eq:D1n}--\ref{eq:D10n}) follows the procedure
of Young, Hao, and Martinsson~\cite[Section 5.3]{Youn12}. The
underlying idea -- to expand the Green's function for the Laplacian in
toroidal harmonics -- is due to Cohl and Tohline~\cite{Cohl99}. The
toroidal harmonics can be evaluated via a recursion whose forward form
is
\begin{equation}
\mathfrak{Q}_{n-\frac{1}{2}}(\chi)=
 \frac{4n-4}{2n-1}\chi\mathfrak{Q}_{n-\frac{3}{2}}(\chi)
-\frac{2n-3}{2n-1}\mathfrak{Q}_{n-\frac{5}{2}}(\chi)\,,
\qquad n=2,\ldots,N\,.
\label{eq:forw}
\end{equation}
Appendix~\ref{app:toro} contains the {\sc Matlab} function {\tt
  toroharm} which evaluates the functions
$\mathfrak{Q}_{-\frac{1}{2}}(\chi)$ and
$\mathfrak{Q}_{\frac{1}{2}}(\chi)$, needed to
initiate~(\ref{eq:forw}). The functions $\mathfrak{R}_n(\chi)$
of~(\ref{eq:Rfrak}) are finite at $\chi=1$, but have logarithmic
singularities in their first (right) derivatives.

\section{Nyström Discretization and kernel evaluation}
\label{sec:discrete}

Our Nyström discretization scheme along $\gamma$ for the modal
equations~(\ref{eq:comp12F}) and~(\ref{eq:comp13F}) and for the
normalization integral of Appendix~\ref{app:A} is, essentially,
identical to the scheme developed in~\cite{HelsKarl14} in a pure
Helmholtz setting. This section only gives a brief review. The scheme
relies on an underlying panel-based 16-point Gauss--Legendre
quadrature with a mesh of $n_{\rm pan}$ quadrature panels on $\gamma$.
The $16n_{\rm pan}$ discretization points play the role of both target
points $r_i$ and source points $r_j$. The underlying quadrature is
used in a conventional way when kernels $K_\alpha(\vec r,\vec r')$ are
sufficiently smooth for $K_{\alpha n}(r_i,r_j)$ to be evaluated by
FFT. For $r_i\approx r_j$, and when convolution is used for $K_{\alpha
  n}(r_i,r_j)$, an explicit kernel-split special quadrature is
activated. Analytical information about the (near) singularities in
$K_{\alpha n}(r,r')$ is exploited in the construction of 16th order
accurate weight corrections, computed on the fly. As to some extent
compensate for the loss of convergence order that comes with the
special quadrature, a procedure of temporary mesh refinment is
adopted. See~\cite{HelsKarl14}, and also~\cite{HelsHols14}, for more
information on these constructions and procedures.

\subsection{The MFIE system and the decomposed magnetic field}

It is worth emphasizing that all $K_{\alpha n}(r,r')$ are singular at
$r=r'$ and that the singularities are inherited by the corresponding
$D_{\alpha n}(r,r')$ of~(\ref{eq:D1n}--\ref{eq:D10n}). The
coefficients $D_{\alpha n}(r,r')$, $\alpha=1,2,3,4$, exhibit
logarithmic singularities as $\gamma\ni r'\to r\in\gamma$. The
coefficients $D_{\alpha n}(r,r')$, $\alpha=5,8,9$, exhibit logarithmic
singularities as $\mathbb{R}^{2+}\setminus\gamma\ni r\to r'\in\gamma$.
The coefficients $D_{\alpha n}(r,r')$, $\alpha=6,7,10$, generally
exhibit logarithmic and Cauchy-type singularities as
$\mathbb{R}^{2+}\setminus\gamma\ni r\to r'\in\gamma$. The quadratures
constructed in~\cite{HelsHols14,HelsKarl14} cover all these
situations.

The appearance of the closed-form expressions for $D_{\alpha n}(r,r')$
may seem somewhat intimidating at first glance. Nevertheless, the
expressions are favorable from a computational perspective. The same
toroidal functions repeat themselves, or occur in combination with
smooth and simple functions independent of $n$. Remember that the
$D_{\alpha n}(r_i,r_j)$ are convolved with parts of $P_n(r_i,r_j)$ to
obtain $K_{\alpha n}(r_i,r_j)$ for $r_i\approx r_j$. Once
$\mathfrak{Q}_{n-\frac{1}{2}}(\chi_{ij})$,
$\mathfrak{R}_n(\chi_{ij})$, and $P_n(r_i,r_j)$, $n=0,\ldots,N$, are
evaluated and the convolution of
$\mathfrak{Q}_{n-\frac{1}{2}}(\chi_{ij})$,
$n\mathfrak{Q}_{n-\frac{1}{2}}(\chi_{ij})$, and
$\mathfrak{R}_n(\chi_{ij})$ with parts of $P_n(r_i,r_j)$ is performed
at all necessary combinations of $r_i$ and $r_j$, the evaluation of
$K_{\alpha n}(r_i,r_j)$, $\alpha=1,2,3,4$, $r_i\approx r_j$, is very
cheap. The evaluation of $K_{\alpha n}(r_i,r_j)$, $\alpha=5,\ldots,10$
requires a few more function evaluations and convolutions.

\subsection{The normalization integral}

Appendix~\ref{app:A} uses the scaled electric scalar potential
$\Psi(\vec r)$, the magnetic vector potential $\vec\Lambda(\vec r)$,
and the surface charge density $\varrho_{\rm s}(\vec r)$. These
quantities are related to $\vec J_{\rm s}(\vec r)$ via
\begin{align}
\Psi(\vec r)&=S_\varsigma\varrho_{\rm s}(\vec r)\,,
\qquad\vec r\in \mathbb{R}^3\,,
\label{eq:Psidef}\\
\vec\Lambda(\vec r)&=S_\varsigma\vec J_{\rm s}(\vec r)\,,
\qquad\vec r\in \mathbb{R}^3\,,
\label{eq:Adef}\\
\varrho_{\rm s}(\vec r)&=
-\frac{\rm i}{k}\nabla_{\rm s}\cdot\vec J_{\rm s}(\vec r),
\qquad\vec r\in\Gamma\,,
\label{eq:scud}
\end{align}
where $\nabla_{\rm s}\cdot(\;)$ is the surface divergence and
$S_\varsigma$ is a single-layer type operator of the form
\begin{equation}
S_{\alpha}g({\vec r})=
\int_\Gamma S_{\alpha}(\vec r,\vec r')g({\vec r'})\,{\rm d}\Gamma'
=\int_\Gamma Z_{\alpha}(\vec r,\vec r')
e^{{\rm i}k\vert\vec r-\vec r'\vert}g({\vec r'})\,{\rm d}\Gamma'\,,
\label{eq:Sdef}
\end{equation}
with static kernel
\begin{equation}
Z_\varsigma(\vec r,\vec r')=\Phi_0(\vec r,\vec r') \quad
{\rm and} \quad 
Z_{\varsigma n}(r,r')=\eta\mathfrak{Q}_{n-\frac{1}{2}}(\chi)\,.
\end{equation}

Azimuthal Fourier coefficients of $\Psi(\vec r)$, $\vec\Lambda(\vec
r)$, and their normal- and tangetial derivatives need to be evaluated
at discretization points $r_j$ along $\gamma$. This, in turn, requires
the introduction and discretization of several new and similar
integral operators of the double-layer type~(\ref{eq:Kdef}) and of the
single-layer type~(\ref{eq:Sdef}). It would, perhaps, carry too far to
explicitly write up closed-form expressions for all $D_\alpha(\vec
r,\vec r')$, $Z_{\alpha}(\vec r,\vec r')$, $D_{\alpha n}(r,r')$, and
$Z_{\alpha n}(r,r')$ involved. The closed-form Fourier coefficients
are derived using the same techniques as in
Section~\ref{sec:azimuthn}. It is important to note that we avoid
using~(\ref{eq:scud}) as a computational formula. Numerical
differentiation leads to loss of precision and also to loss of
convergence order in a panel-based setting. Rather, the surface charge
density is obtained from the solution to a Fredholm second kind
integral equation, as recommended in~\cite{Vicoetal13}. We also take
advantage of two useful relations between $\vec\Lambda(\vec r)$ and
derivatives of $\Psi(\vec r)$. Appendix~\ref{app:B} provides some
detail.

\begin{figure}[t!]
  \centering 
\includegraphics[height=101mm]{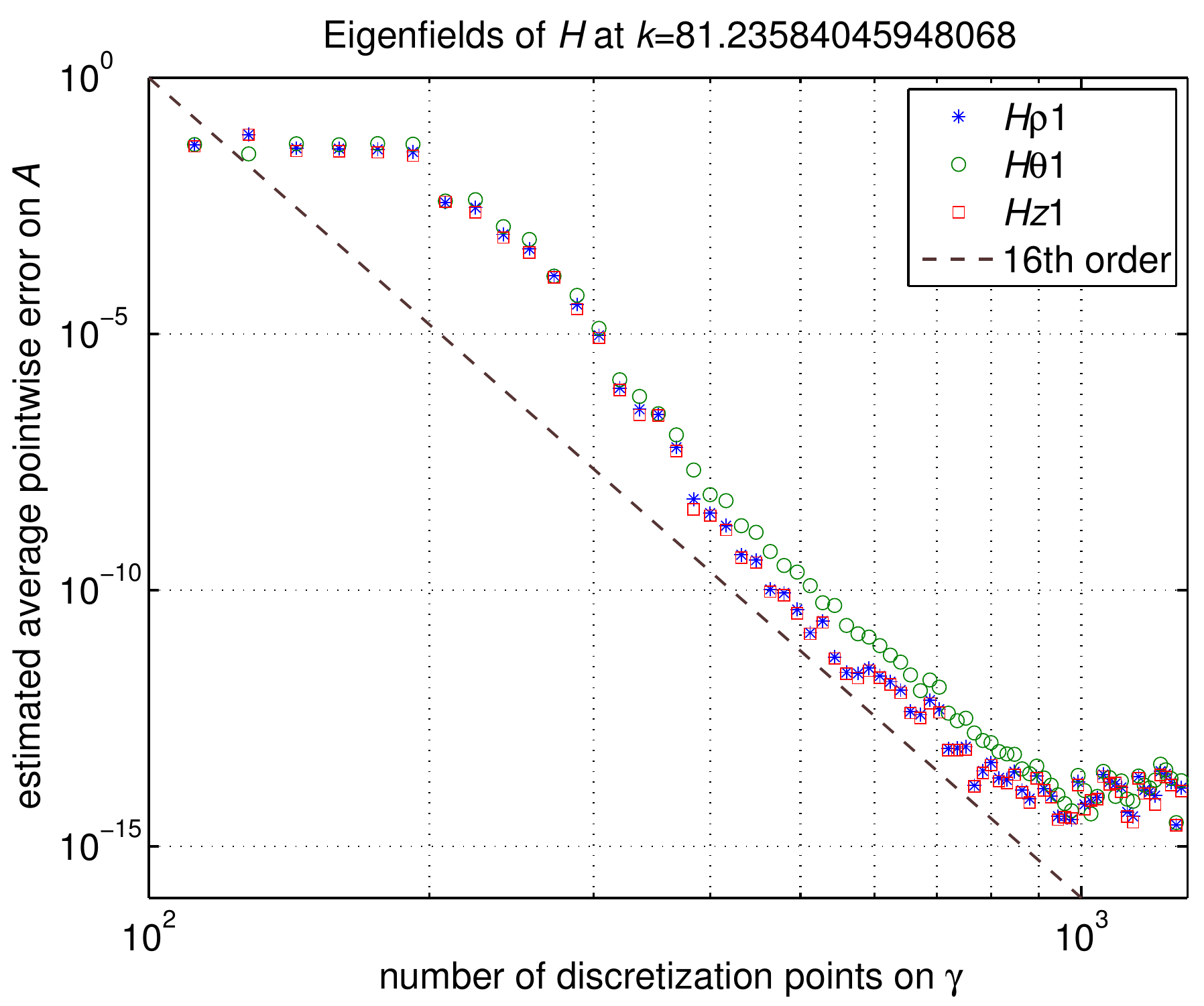}
\caption{\sf Convergence of the coefficients $H_{\rho 1}(r)$, 
  $H_{\theta 1}(r)$, and $H_{z 1}(r)$. The average accuracy has
  converged to between 13 and 14 digits at $864$ discretization points
  on $\gamma$, corresponding to about 16 points per wavelength along
  $\gamma$.}
\label{fig:conv81}
\end{figure}

\section{Numerical examples}
\label{sec:numerical}

Our Fourier--Nyström scheme is implemented in {\sc Matlab}. The code
was first verified by comparison with analytic solutions for the
special case of $\Gamma$ being the unit sphere~\cite[Chapter
9]{Stratton1941}. Eigenwavenumbers $k$ corresponding to a few
wavelengths across $V$ gave coefficients $J_{\tau n}(r),J_{\theta
  n}(r)$, $r\in\gamma$, with relative $L^2$-errors of about $10^{-14}$
and coefficients $H_{\rho n}(r),H_{\theta n}(r),H_{zn}(r)$, $r\in A$,
with pointwise errors of, typically, the same magnitude.
Eigenwavenumbers corresponding to 32 wavelengths across $V$ gave
$J_{\tau n}(r),J_{\theta n}(r)$ with $L^2$-errors of about $10^{-13}$
and $H_{\rho n}(r),H_{\theta n}(r),H_{zn}(r)$ with pointwise errors
ranging from $10^{-16}$ to $10^{-11}$. The largest errors occurred for
$r$ close to $\gamma$ in connection with high azimuthal indices $n$.
We also compared evaluations of coefficients for the cavity with the
star-shaped cross-section shown in Figure~\ref{fig:geometry}(a) with
results obtained from COMSOL Multiphysics. Eigenwavenumbers
corresponding to about two wavelengths across $V$ gave results which
agreed to all significant digits that COMSOL Multiphysics could
produce.

We now present two more detailed numerical examples for the normalized
magnetic eigenfields of the body of revolution in
Figure~\ref{fig:geometry}(a). The purpose is to confirm that our
solver meets many of the requirements imposed on a wakefield solver.
The generating curve $\gamma$ is parameterized as
\begin{equation}
r(t)=(\rho(t),z(t))
=(1+0.25\cos(5t))(\sin(t),\cos(t))\,,\qquad 0\leq t\leq\pi\,,
\label{eq:star}
\end{equation}
which is the same curve that was used in the examples of
\cite{HelsKarl14}. The integer $N$, controlling FFT operations, is
chosen as $N=\max\{120,4n_{\rm pan}+n\}$. The {\sc Matlab} code is
executed on a workstation equipped with an Intel Core i7 CPU at 3.20
GHz and 64 GB of memory.

\begin{figure}[t!]
  \centering \includegraphics[height=50mm]{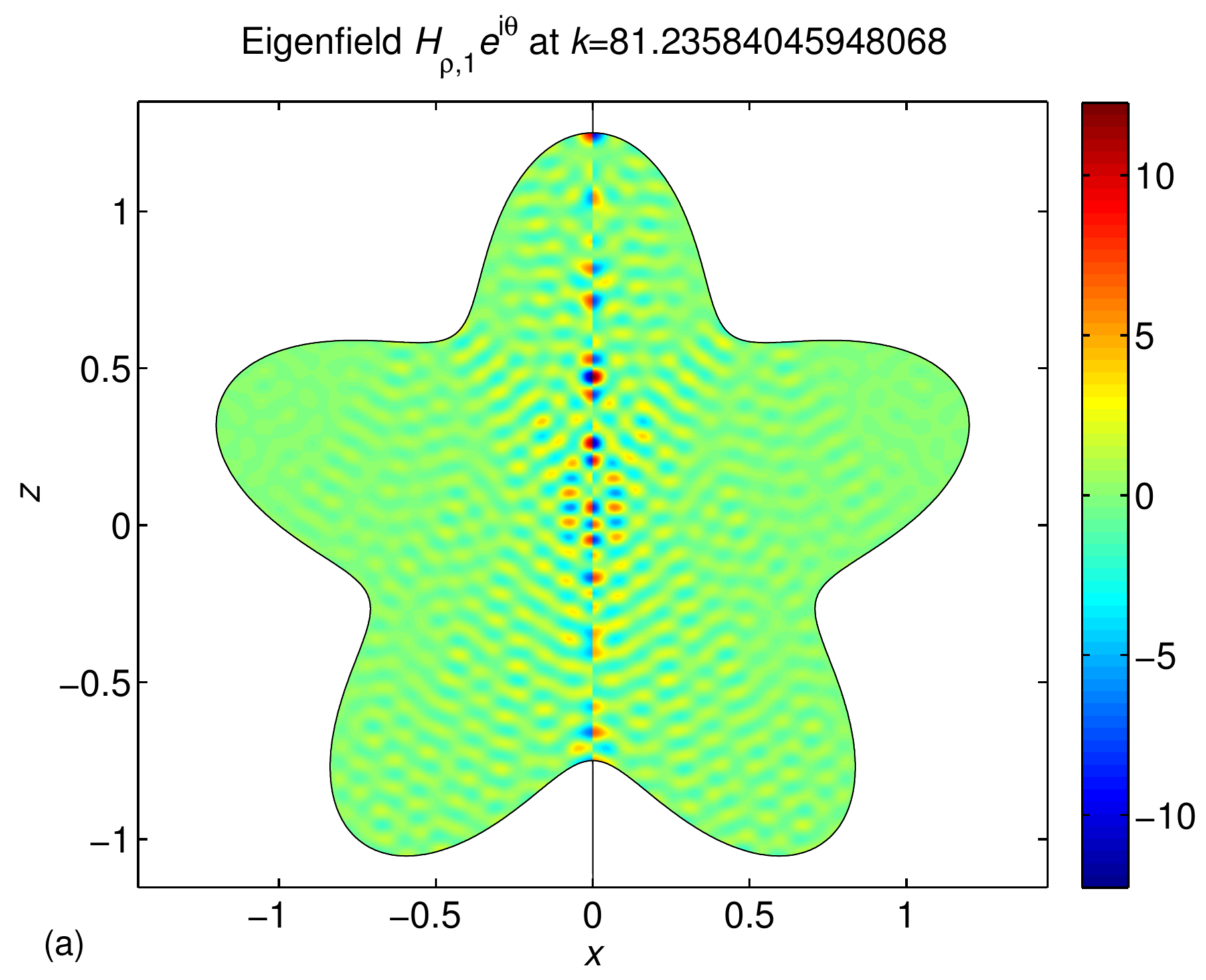}
  \includegraphics[height=50mm]{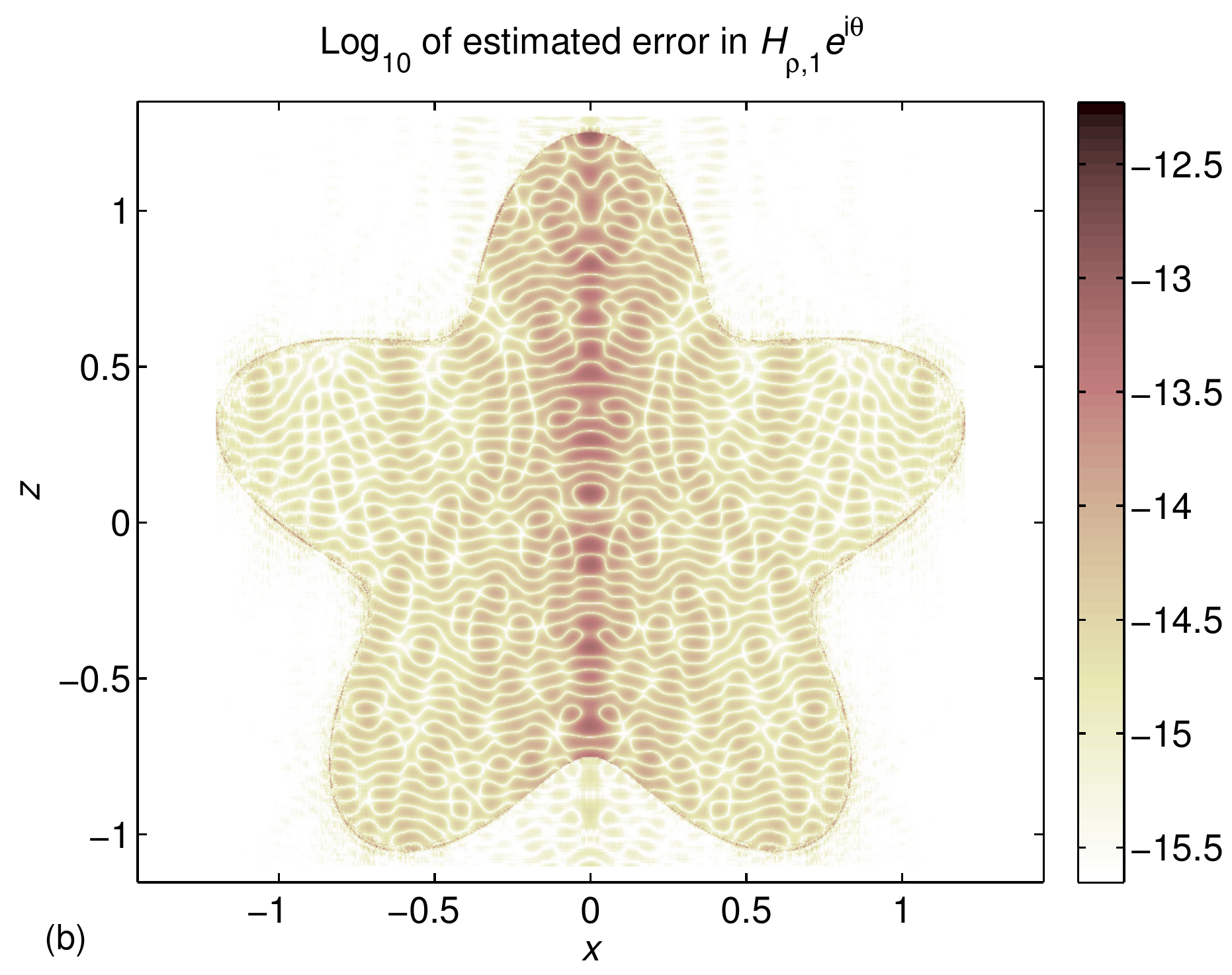}
  \includegraphics[height=50mm]{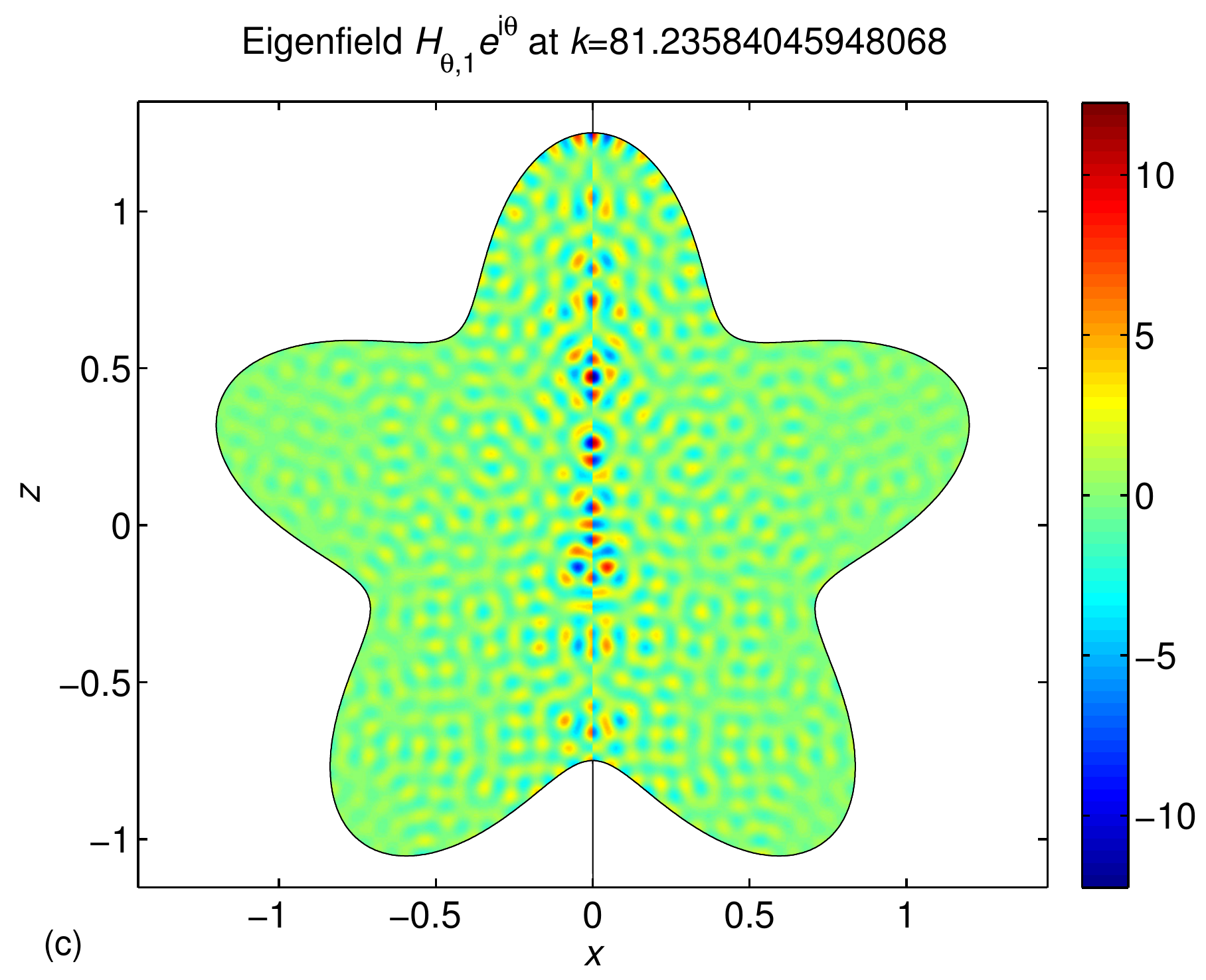}
  \includegraphics[height=50mm]{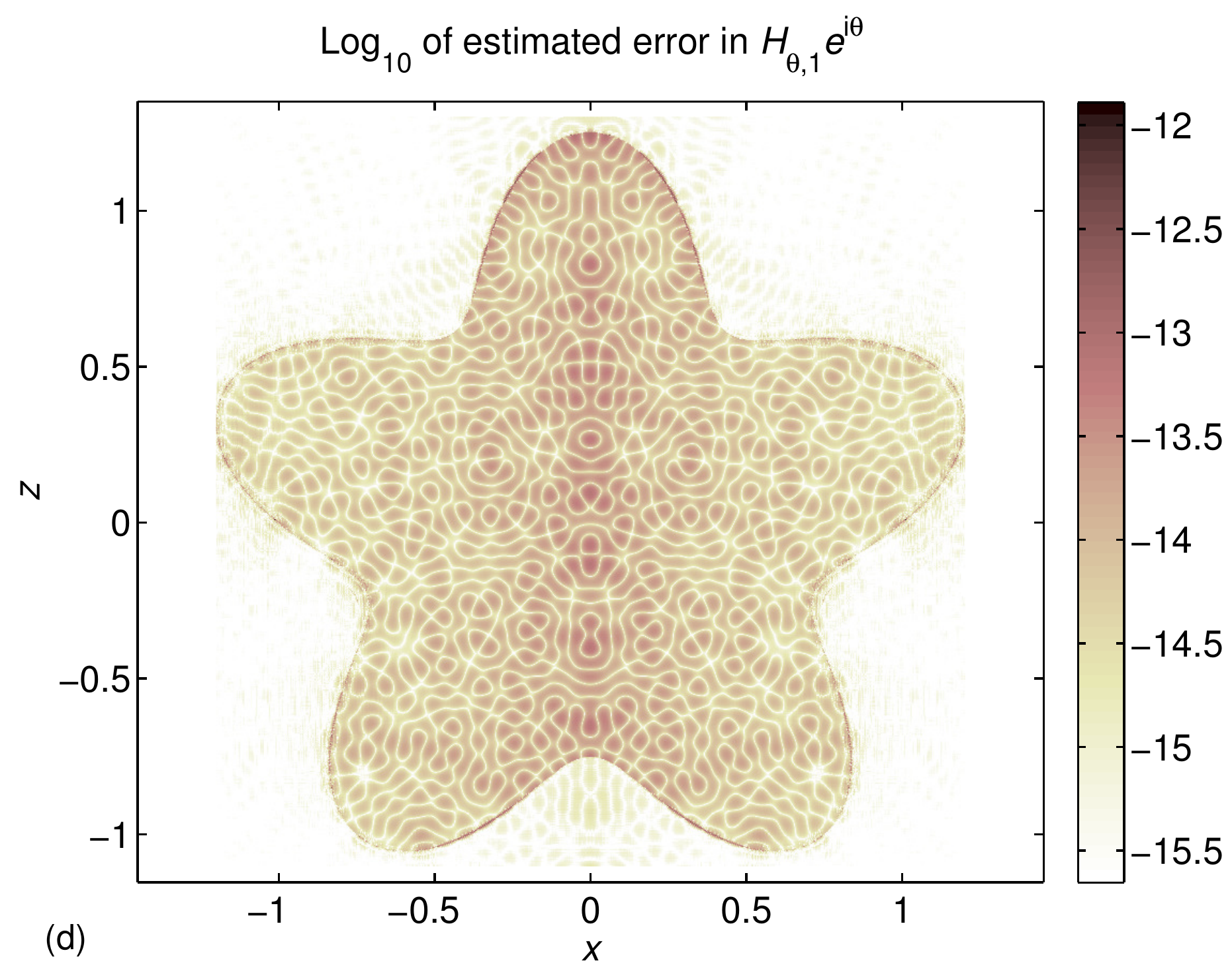}
  \includegraphics[height=50mm]{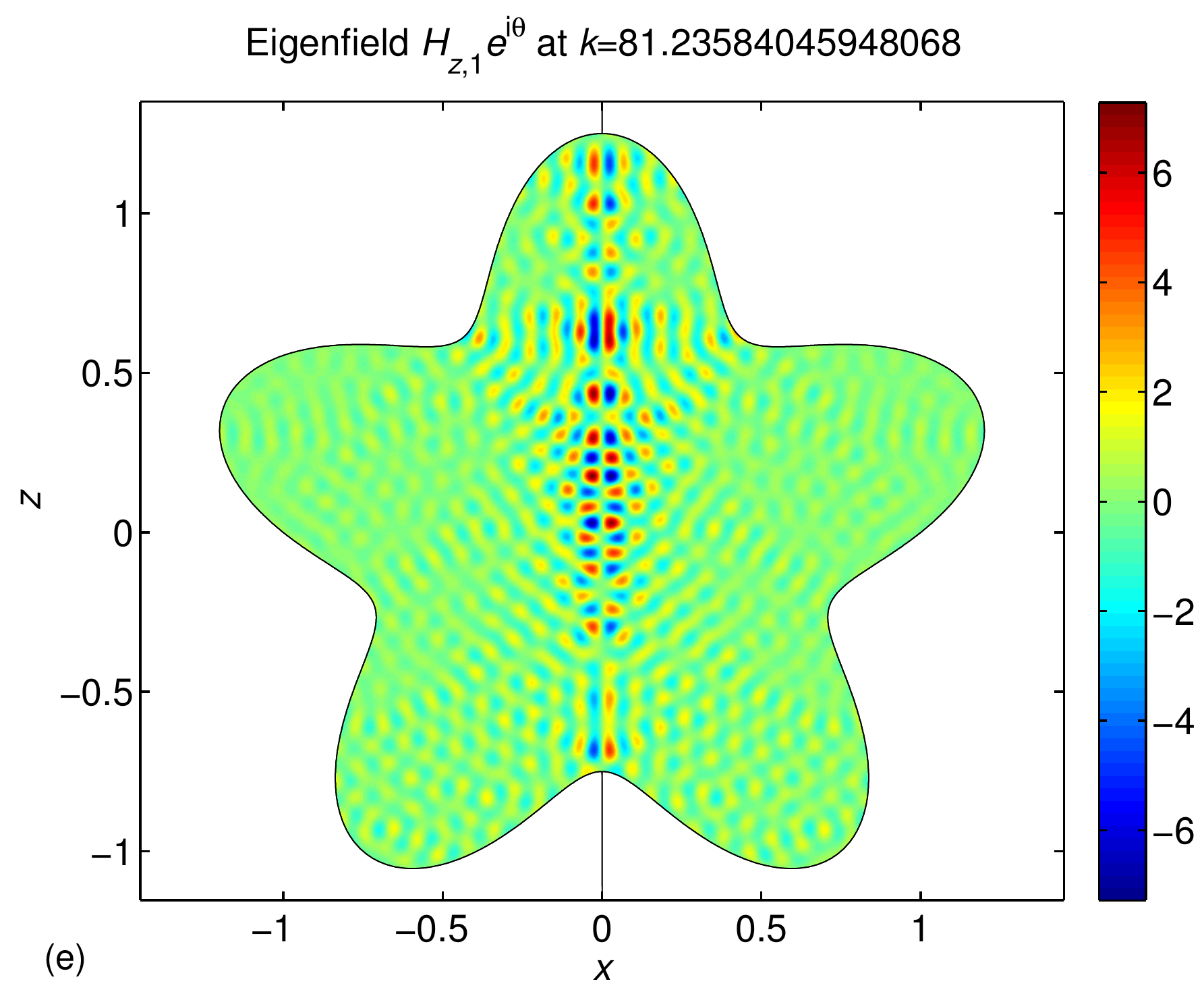}
  \includegraphics[height=50mm]{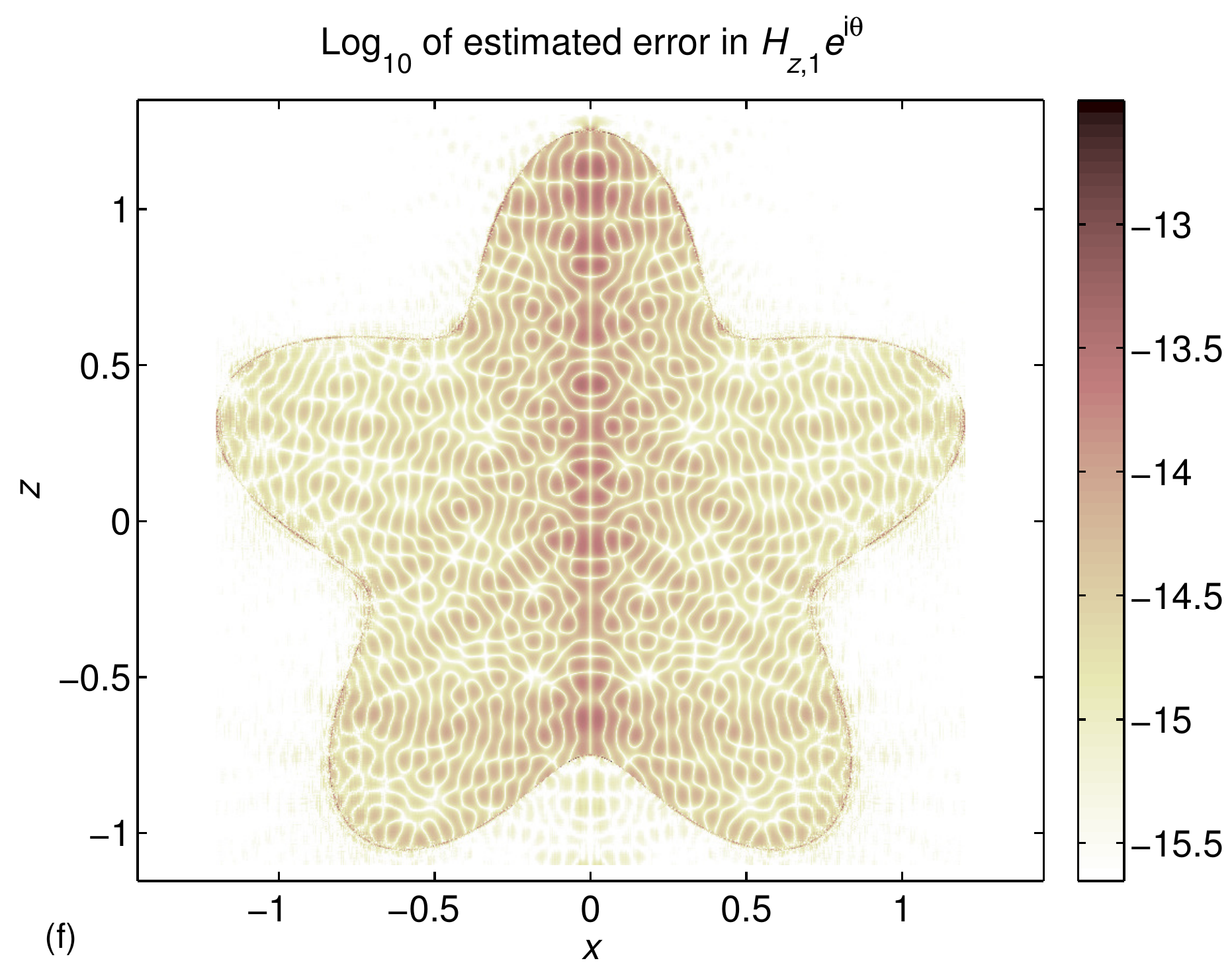}
\caption{\sf Left: (a), (c), (e) show $H_{\rho 1}(r)e^{{\rm
      i}\theta}$, $H_{\theta 1}(r)e^{{\rm i}\theta}$ and
  $H_{z1}(r)e^{{\rm i}\theta}$ at $k=81.23584045948068$ and for
  $\theta=0$ and $\theta=\pi$. Right: (b), (d), (f) show $\log_{10}$
  of the estimated pointwise error with $1136$ discretization points
  on $\gamma$.}
\label{fig:n1}
\end{figure}

\begin{figure}[t!]
\centering
\includegraphics[height=50mm]{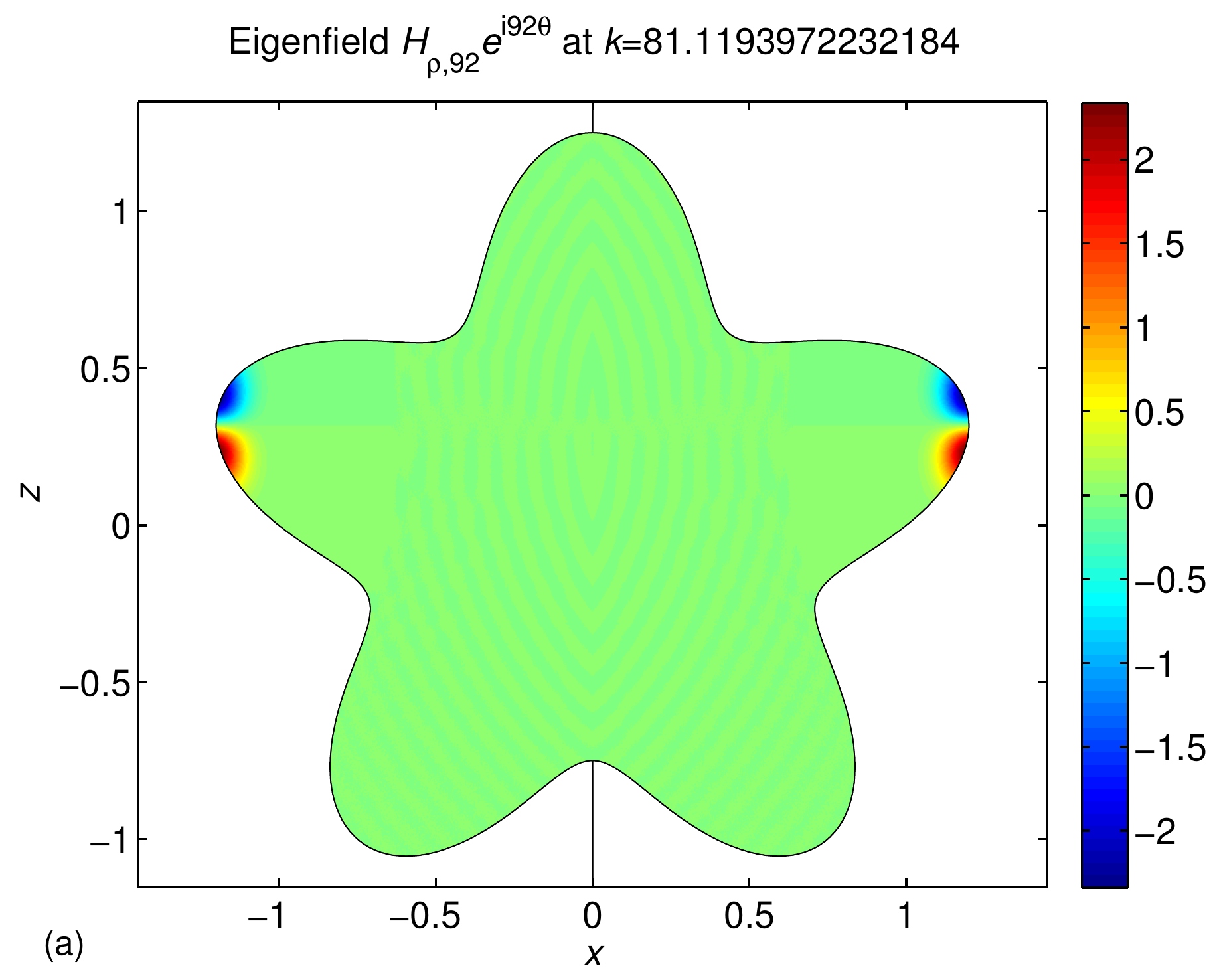}
\includegraphics[height=50mm]{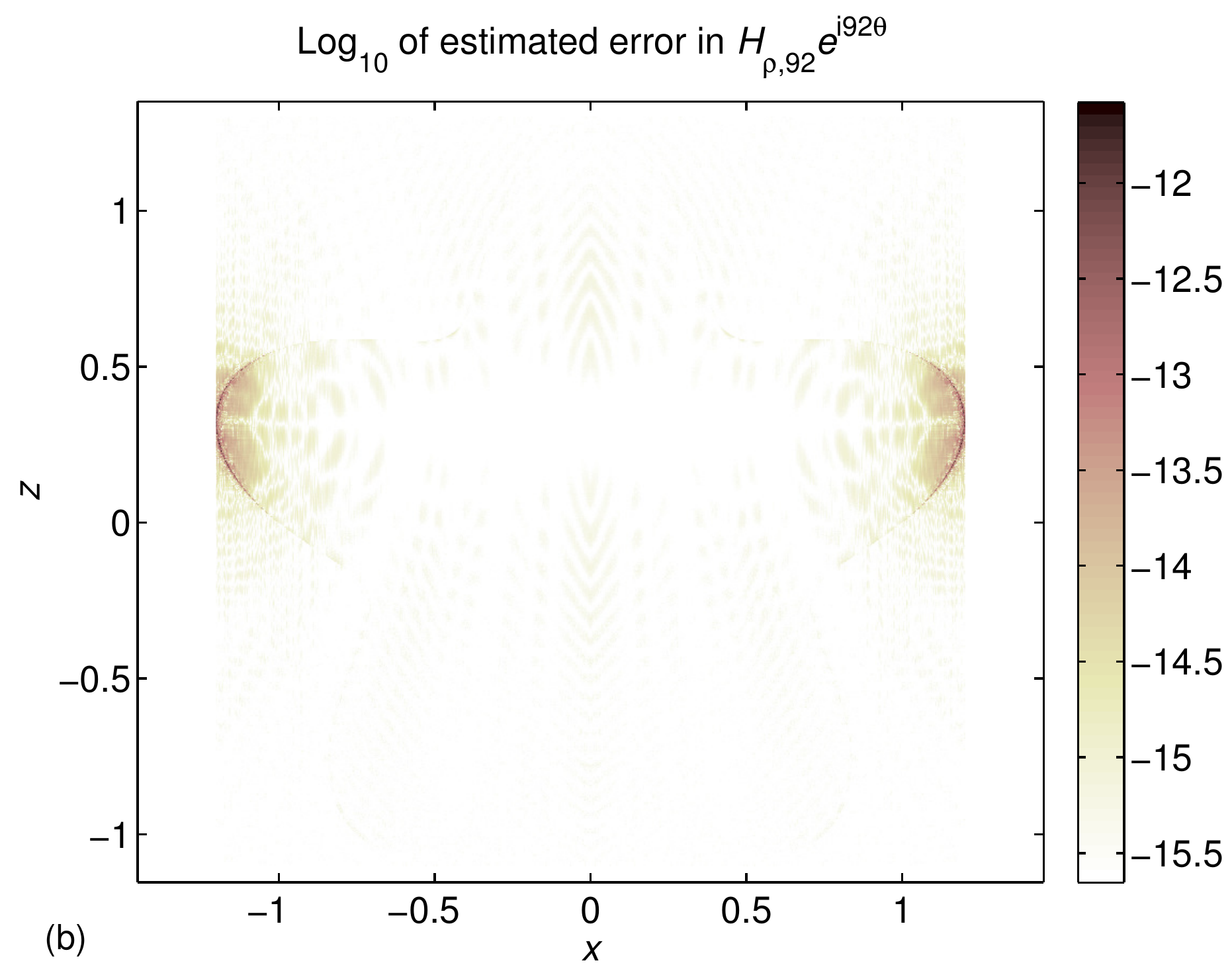}
\includegraphics[height=50mm]{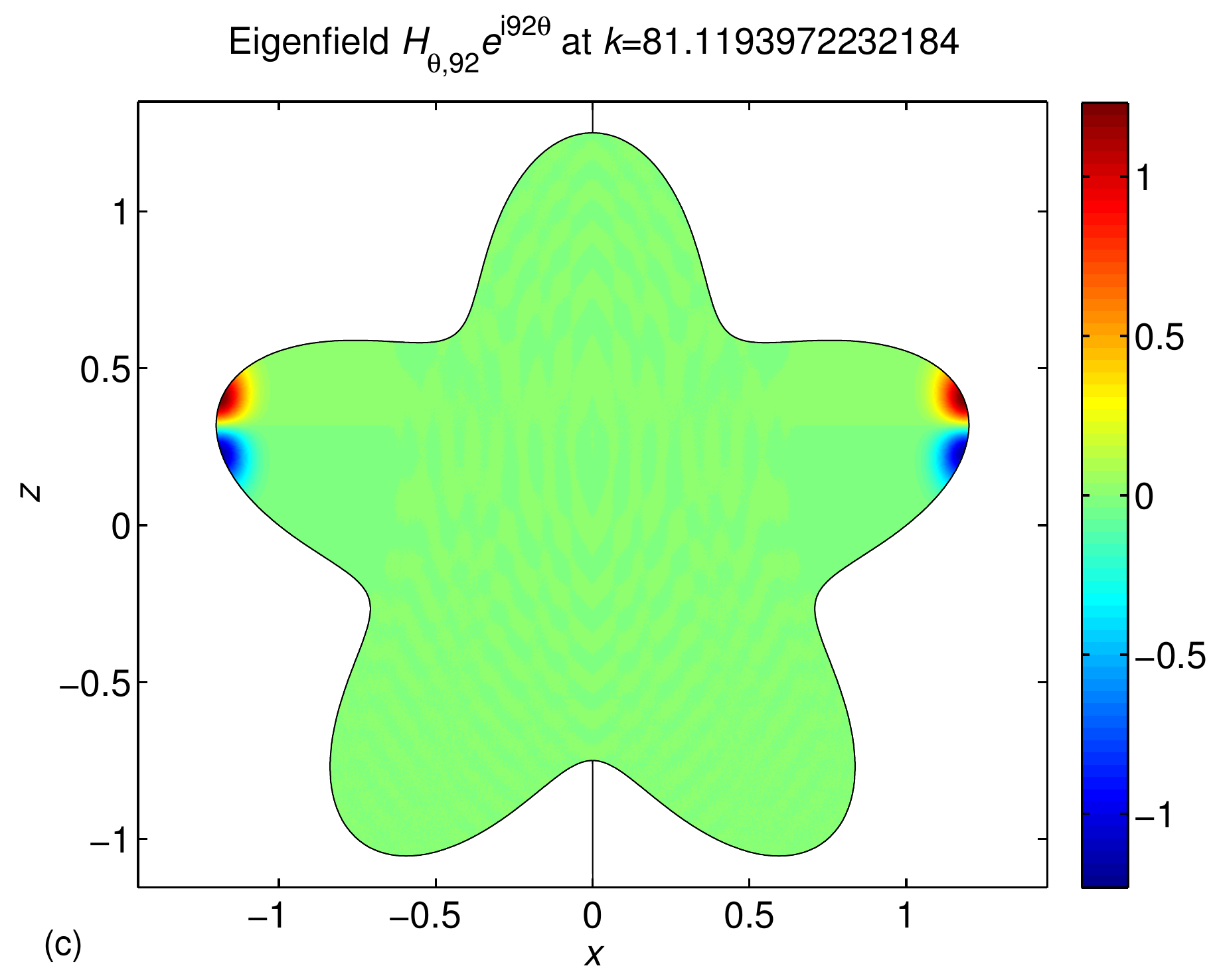}
\includegraphics[height=50mm]{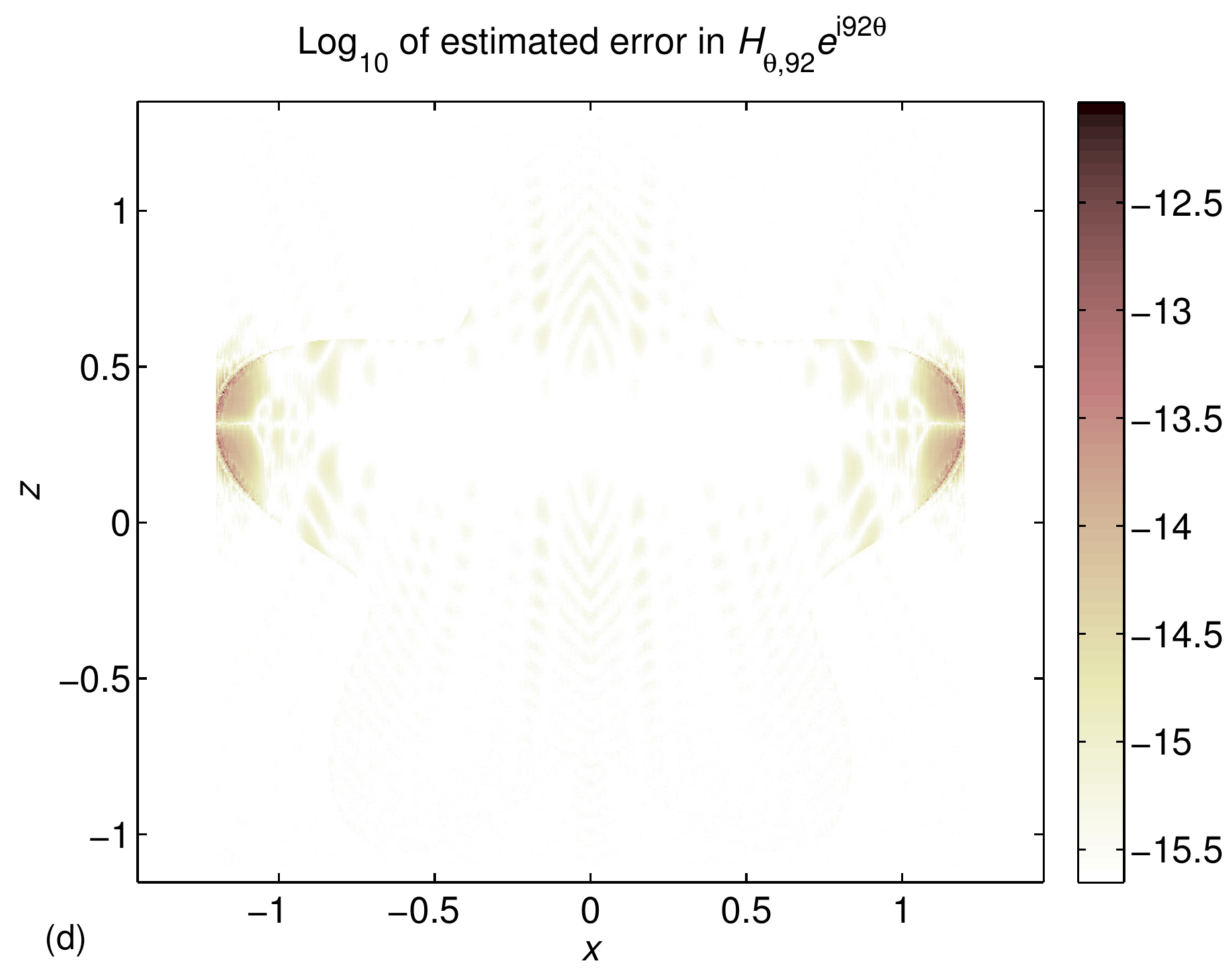}
\includegraphics[height=50mm]{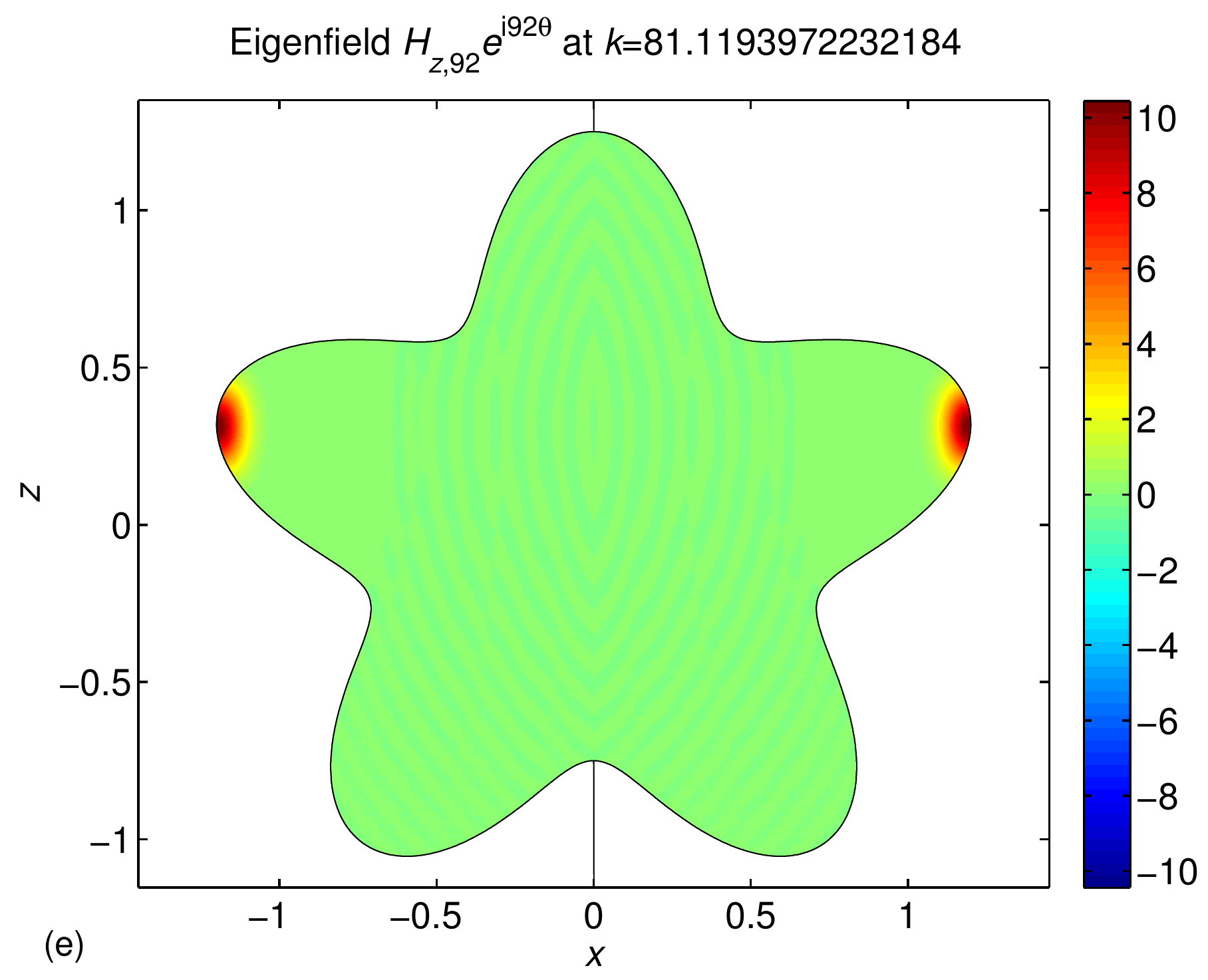}
\includegraphics[height=50mm]{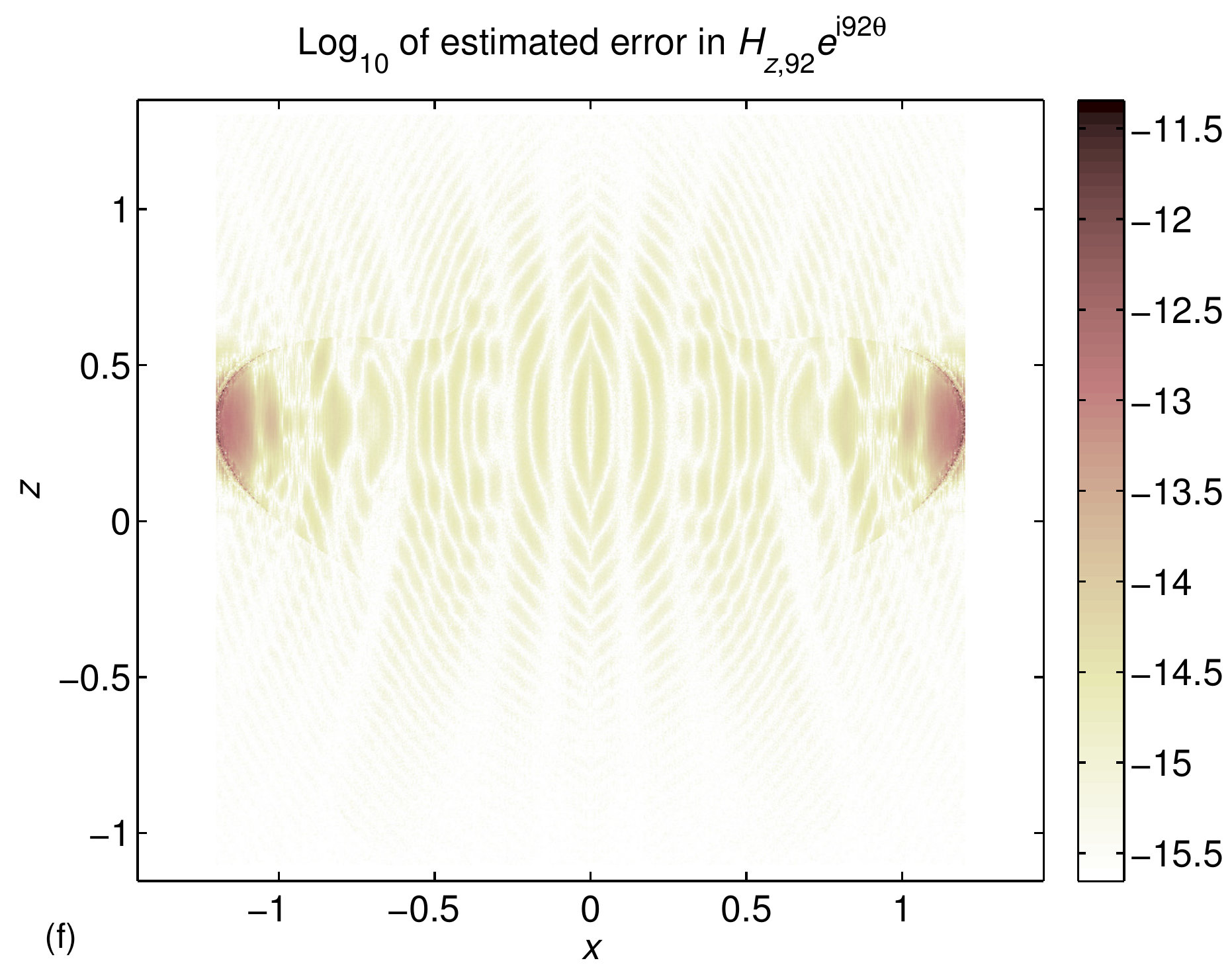}
\caption{\sf Left: (a), (c), (e) show $H_{\rho 92}(r)e^{{\rm i}92\theta}$, 
  $H_{\theta 92}(r)e^{{\rm i}92\theta}$ and $H_{z92}(r)e^{{\rm
      i}92\theta}$ at $k=81.1193972232184$ and for $\theta=0$ and
  $\theta=\pi$. Right: (b), (d), (f) show $\log_{10}$ of the estimated
  pointwise error with $1024$ discretization points on $\gamma$.}
\label{fig:n96}
\end{figure}

The first example concerns an eigenfield with $k=81.23584045948068$,
corresponding to a generalized diameter of $V$ of about 31
wavelengths, and $n=1$. The MFIE is solved repeatedly on an
increasingly refined uniform mesh. The coefficient vector $(H_{\rho
  1}(r),H_{\theta 1}(r),H_{z 1}(r))$ is evaluated at $25138$ field
points $r$, placed on a Cartesian grid in $A$. Figure~\ref{fig:conv81}
shows that our solver exhibits 16th order convergence, as expected.
The {\it pointwise error} refers to an estimated absolute pointwise
error of $H_{\alpha n}(r)$, $\alpha=\rho,\theta,z$, divided with the
largest value of $(|H_{\rho n}(r)|^2+|H_{\theta n}(r)|^2+|H_{z
  n}(r)|^2)^{\frac{1}{2}}$, $r\in A$. The estimated absolute error in
$H_{\alpha n}(r)$ is, in turn, taken as the difference between
$H_{\alpha n}(r)$ and a reference solution obtained with a mesh
containing 50 per cent more quadrature panels and a larger $N$.
Figure~\ref{fig:conv81} also shows that the average pointwise
precision in $A$ has converged to ten digits already at about ten
points per wavelength along $\gamma$ and that it saturates at between
13 and 14 digits. This, rather high, achievable accuracy is only one
digit worse than what is reported for the similar, but simpler,
acoustic problem in~\cite{HelsKarl14} at $k\approx 19$.

Figure~\ref{fig:n1} shows $H_{\alpha 1}(r)e^{{\rm i}\theta}$,
$\alpha=\rho,\theta,z$, and $\theta=0,\pi$, at $490000$ field points
on a Cartesian grid in the square $x=[-1.2,1.2]$ and $z=[-1.1,1.3]$
along with estimated pointwise errors. The experiment uses $n_{\rm
  pan}=71$ in order to assure that the convergence has saturated. The
reference solution outside the cavity is the null field $H_{\alpha
  1}(r)=0$, compare~(\ref{eq:Hrep}). The complicated standing wave
patterns in the $xz$-plane, visible in the left images, are typical
for eigenfields with large $k$ in combination with small $n$.
Electric and magnetic eigenfields with $n=0$ and $n=1$ are non-zero on
the symmetry axis in axially symmetric cavities. Since the beam of
particles in accelerators travels along the symmetry axis, it
interacts strongly with these fields. Eigenfields with $n>1$ are zero
on the symmetry axis and are less important in accelerator technology.
It is interesting to observe that the field errors in
Figure~\ref{fig:n1}, right images, are comparable to those reported by
Barnett~\cite{Barn14} for planar Dirichlet Helmholtz problems exterior
to objects with diameters ranging from $12$ to $100$ wavelengths.

The second example concerns and eigenfield with $k=81.1193972232184$
and $n=92$. This happens to be the eigenfield with the largest $n$ for
$k<82$. The eigenfields with the largest $n$ for a given $k$ are
confined to the parts of the cavity that are farthest from the
symmetry axis, as seen in Figure \ref{fig:n96}. Such fields are called
whispering-gallery modes (WGM). They are of very little interest for
particle accelerators but have recently become important in
nano-optics. See~\cite{Tamboli07} for interesting applications of WGM.
Figure \ref{fig:n96} depicts the magnetic eigenfields and the
corresponding errors. The convergence with mesh refinement (not shown)
is similar to that of Figure~\ref{fig:conv81}. This is in accordance
with our experience that the convergence speed and achievable accuracy
of our solver are, more or less, insensitive to $n$. The WGM are
interesting objects to analyze numerically.

\section{Conclusion and outlook}
\label{sec:conclus}

We have presented a competitive numerical solver for the determination
of normalized magnetic eigenfields in axially symmetric microwave
cavities with smooth surfaces. It is based on the following key
elements: the magnetic field integral equation, a surface integral for
normalization, a Fredholm second kind integral equation for the
surface charge density, and a high-order convergent Fourier--Nyström
discretization scheme. In the near future, we will extend the solver
so that it can determine electric eigenfields. We may also explore the
whispering gallery modes of structures that better resemble those used
in nano-optics.

A more challenging task, that we plan to take on, is to extend our
solver so that it can handle non-smooth surfaces. Starting in 2008,
our group has developed an extremely robust technique for the accurate
solution of integral equations on piecewise smooth surfaces -- most
recently applied to planar scattering problems~\cite{HelsKarl13} and
to electrostatic problems in $\mathbb{R}^3$~\cite{HelsP13}. If applied
to the time harmonic Maxwell equations, this technique will enable the
accurate determination of eigenfields, and by that the prediction of
wakefields, in most types of cavities and flanges used in particle
accelerators.

\section*{Acknowledgement}

This work was supported by the Swedish Research Council under contract
621-2011-5516.

\appendix

\section{The normalization integral as a surface integral}
\label{app:A}

It is convenient, and sometimes necessary, to normalize magnetic
eigenfields 
\begin{equation}
\vec H(\vec r)=\frac{\vec H_n(r)}{\sqrt{2\pi}}e^{{\rm i}n\theta}
\label{eq:Heig}
\end{equation}
according to~(\ref{eq:Hnorm}), that is, such that $\|\vec H\|=1$.
In~(\ref{eq:Heig}) we have introduced
\begin{equation}
\vec H_n(r)=\vec\rho H_{\rho n}(r)+\vec\theta H_{\theta n}(r)
           +\vec zH_{zn}(r)\,.
\end{equation}
Normalized eigenfields are important in wakefield calculations where
an eigenfield amplitude is determined by integration along the
symmetry axis of the product of the electric eigenfield with a beam
current. The normalization integral in~(\ref{eq:Hnorm}) is also needed
for the extraction of $Q$-values.

This appendix presents an expression for the normalization integral
that is cheaper to evaluate than the volume integral
in~(\ref{eq:Hnorm}). In fact, the normalization integral will be
expressed as a line integral over $\gamma$ involving readily
accessible azimuthal Fourier coefficients. To make~(\ref{eq:Hnorm})
hold within our numerical scheme, the coefficients $J_{\tau n}(r)$ and
$J_{\theta n}(r)$ obtained from~(\ref{eq:comp12F}) are normalized with
the value of $\|\vec H\|$ according to the formulas below. Then the
modal representation~(\ref{eq:comp13F}) is automatically consistent
with~(\ref{eq:Hnorm}).

The scaled electric scalar potential $\Psi(\vec r)$
of~(\ref{eq:Psidef}) and the magnetic vector potential
$\vec\Lambda(\vec r)$ of~(\ref{eq:Adef}) are related by the Lorenz
gauge condition
\begin{equation}
{\rm i}k\Psi=\nabla\cdot\vec \Lambda\,.
\label{lorenz}
\end{equation}
In $\mathbb{R}^3$ the potentials satisfy
\begin{align}
\nabla^2\Psi&=-k^2\Psi\,,\\
\nabla^2\vec\Lambda&=-k^2\vec\Lambda\,,\\
\vec H&=\nabla\times\vec\Lambda\,,
\label{eq:HfromL}
\end{align}
and it is easy to show
\begin{equation}
|\nabla\times\vec\Lambda|^2
=\nabla\cdot(\vec\Lambda^\ast\times(\nabla\times\vec\Lambda))
+{\rm i}k\nabla\cdot(\vec\Lambda^\ast\Psi)
-k^2|\Psi|^2+k^2|\vec\Lambda|^2\,.
\label{eq:cross}
\end{equation}
Let $g(\vec r)$ be a Laplace eigenfunction such that
$\nabla^2g=-k^2g$. Then
\begin{align}
\nabla\cdot(\vec r|g|^2)&=3|g|^2+2\Re{\rm e}\{g^\ast(\vec r\cdot\nabla)g\}\,,
\nonumber\\
\nabla\cdot(\vec r|\nabla g|^2)&=
3|\nabla g|^2+2\Re{\rm e}\{(\nabla g^\ast\cdot\nabla)\nabla g\cdot\vec r\}\,,
\nonumber\\
\Re{\rm e}\{\nabla\cdot((\vec r\cdot\nabla)g^\ast\nabla g)\}&=
|\nabla g|^2-k^2 \Re{\rm e}\{g^\ast(\vec r\cdot\nabla)g\}
+\Re{\rm e}\{(\nabla g^\ast\cdot\nabla)\nabla g\cdot\vec r\}\,,
\nonumber
\end{align}
and, by that,
\begin{equation}
|g|^2=\frac{1}{2k^2}
\nabla\cdot\left(\vec r\left(k^2|g|^2-|\nabla g|^2\right)
+\Re{\rm e}\{\nabla g^\ast(2(\vec r\cdot\nabla)+1)g\}\right)\,.
\label{eq:Barnett1}
\end{equation}

Now, from~(\ref{eq:Hnormdef}),~(\ref{eq:HfromL}),~(\ref{eq:cross}) and
Gauss' theorem one obtains
\begin{equation}
\|\vec H\|^2=k^2\left(\|\vec\Lambda\|^2-\|\Psi\|^2\right)
+\int_\Gamma\vec\Lambda^\ast\cdot(\vec J_{\rm s}
+{\rm i}k\vec\nu\Psi)\,{\rm d}\Gamma\,.
\label{eq:magne}
\end{equation}
The relation \eqref{eq:Barnett1} with $g(\vec r)$ first equal to
$\Psi(\vec r)$ and then to each of the Cartesian components of
$\vec\Lambda(\vec r)$, together with~(\ref{eq:Hnormdef}) and Gauss'
theorem, convert the squared norms on the right hand side of
\eqref{eq:magne} to the surface integrals
\begin{align}
\|\Psi\|^2&=\frac{1}{2k^2}\int_\Gamma\vec\nu\cdot\left(\vec r\left(k^2|\Psi|^2 
-|\nabla\Psi|^2\right)
+\Re{\rm e}\{\nabla\Psi^\ast(2(\vec r\cdot\nabla)
+1)\Psi\}\right)\,{\rm d}\Gamma\,,
\label{eq:Psisq}\\
\|\vec\Lambda\|^2&=\frac{1}{2k^2}\int_{\Gamma}
\vec\nu\cdot\left(\vec r\left(k^2|\vec\Lambda|^2-|\nabla\vec\Lambda|^2\right)
+\Re{\rm e}\{\nabla\vec\Lambda^\ast\cdot(2(\vec r\cdot\nabla)
+1)\vec\Lambda\}\right)\,{\rm d}\Gamma\,,
\label{eq:Asq}
\end{align}
where interior limits are to be taken for integrands that are
discontinuous at $\Gamma$. By this, $\|\vec H\|^2$ is expressed as a
surface integral over $\Gamma$. We remark that~(\ref{eq:Psisq}) is
Barnett's relation~\cite[equation~(12)]{Barn06B} generalized to
complex-valued eigenfunctions. See also~\cite[Lemma~3.1]{Barn06A}, for
an $\mathbb{R}^2$ version of~(\ref{eq:Psisq}).

Equation \eqref{eq:magne} with~(\ref{eq:Hnormdef}) and~(\ref{eq:Heig})
can be written in terms of
\begin{align}
\vec\Lambda_n(r)&=\vec\rho\Lambda_{\rho n}(r)
                 +\vec\theta\Lambda_{\theta n}(r)
                 +\vec z\Lambda_{zn}(r)\,,\qquad r\in A\cup\gamma\,,\\
\vec\Lambda_n(r)&=\vec\tau\Lambda_{\tau n}(r)
                 +\vec\theta\Lambda_{\theta n}(r)
                 +\vec\nu\Lambda_{\nu n}(r)\,,\qquad r\in\gamma\,,
\label{eq:Lttn}\\
\vec J_{{\rm s}n}(r)&=\vec\tau J_{\tau n}(r)
                     +\vec\theta J_{\theta n}(r)\,,\qquad r\in\gamma\,,
\end{align}
and $\Psi_n(r)$, $r\in A\cup\gamma$, as
\begin{equation}
\|\vec H\|^2=
k^2\int_A\left(|\vec\Lambda_n|^2-|\Psi_n|^2\right)\rho\,{\rm d}A
+\int_\gamma\left(\vec\Lambda_n^\ast\cdot\vec J_{{\rm s}n}
+{\rm i}k\Lambda_{\nu n}^\ast\Psi_n\right)\rho\,{\rm d}\gamma\,.
\label{eq:magne2}
\end{equation}
Integration over $\theta$, with $\nabla=\vec\tau(\vec
\tau\cdot\nabla)+\vec\theta(\vec \theta\cdot\nabla)+\vec\nu(\vec
\nu\cdot\nabla)$, in~(\ref{eq:Psisq}) and~(\ref{eq:Asq}) gives
\begin{align}
&\int_A|\Psi_n|^2\rho \,{\rm d}A=
-\frac{1}{2k^2}\int_\gamma\frac{\nu\cdot r}{\rho^2}\left(n^2-k^2\rho^2\right)
|\Psi_n|^2\rho\,{\rm d}\gamma
\nonumber\\
&\qquad-\frac{1}{2k^2}\int_\gamma\nu\cdot r
 \left(\left|(\partial_{\vec\tau}\Psi)_n\right|^2
      -\left|(\partial_{\vec\nu^+}\Psi)_n\right|^2\right)
\rho\,{\rm d}\gamma
\nonumber\\
&\qquad+\frac{1}{2k^2}\int_\gamma\Re{\rm e}
\left\{\left(2\tau\cdot r(\partial_{\vec\tau}\Psi^\ast)_n+
\Psi^\ast_n\right)(\partial_{\vec\nu^+}\Psi)_n\right\}
\rho\,{\rm d}\gamma\,,
\label{eq:Psq2}\\
&\int_A|\vec \Lambda_n|^2\rho\,{\rm d}A
=-\frac{1}{2k^2}\int_\gamma\frac{\nu\cdot r}{\rho ^2}\left(
(n^2-k^2\rho^2+1)|\vec\Lambda_n|^2
-|\Lambda_{zn}|^2\right)\rho\,{\rm d}\gamma\nonumber\\
&\qquad-\frac{1}{2k^2}\int_\gamma\nu\cdot r\left(
 \left|(\partial_{\vec\tau}\vec\Lambda)_n\right|^2
-\left|(\partial_{\vec\nu^+}\vec\Lambda)_n\right|^2
-\frac{4n}{\rho^2}\Im{\rm m}\left\{
 \Lambda_{\rho n}^\ast\Lambda_{\theta n}\right\}\right)
 \rho\,{\rm d}\gamma\nonumber\\
&\qquad+\frac{1}{2k^2}\int_\gamma\Re{\rm e}\left\{
(2\tau\cdot r(\partial_{\vec\tau}\vec\Lambda^\ast)_n+\vec\Lambda_n^\ast)
\cdot(\partial_{\vec\nu^+}\vec\Lambda)_n\right\}\rho\,{\rm d}\gamma\,,
\label{eq:Asq2}
\end{align}
where, for directional derivatives of a function $g(\vec r)$, we have
used
\begin{align}
\partial_{\vec\tau}g(\vec r)&=
\vec\tau\cdot\nabla g(\vec r)\,,\qquad \vec r\in\Gamma\,,\\
\partial_{\vec\nu^+}g(\vec r^\circ)&=
\lim_{V\ni \vec r\to \vec r^\circ}
\vec\nu^\circ\cdot\nabla g(\vec r)\,,\qquad \vec r^\circ\in\Gamma\,.
\end{align}
By this, $\|\vec H\|^2$ is expressed as a line integral over $\gamma$.
It remains to relate all terms in the integrands of~(\ref{eq:magne2}),
with~(\ref{eq:Psq2}) and~(\ref{eq:Asq2}), to $J_{\tau n}(r)$ and
$J_{\theta n}(r)$. This is the topic of Appendix~\ref{app:B}.

\section{$\Psi$ and $\vec\Lambda$ and their $\vec\tau$- and 
         $\vec\nu^+$-derivatives}
\label{app:B}

In order to evaluate~(\ref{eq:magne2}) from the solution
to~(\ref{eq:comp12F}), the Fourier coefficients of $\Psi(\vec r)$ and
$\vec\Lambda(\vec r)$, and their derivatives with respect to
$\vec\tau$ and $\vec\nu^+$ need to be related to $J_{\tau n}(r)$ and
$J_{\theta n}(r)$. This process, which is carried out for $\vec H(\vec
r)$ in Section~\ref{sec:magexp} and Section~\ref{sec:Fourier},
consists of three steps: First integral representations in terms of
$\vec J_{\rm s}(\vec r)$ are found for $\Psi(\vec r)$,
$\vec\Lambda(\vec r)$, and their derivatives; then these
representations are expanded in Fourier series; and finally
closed-form expressions are constructed for transformed static
kernels. This appendix provides additional details on how to obtain
the coefficients in~(\ref{eq:magne2}) and gives complete information
for some of them. We also review relations that offer simpler and more
accurate coefficient evaluation in certain situations.

The integral representation of $\vec\Lambda(\vec r)$ in terms of $\vec
J_{\rm s}(\vec r)$ is given by~(\ref{eq:Adef}). For $\vec r\in\Gamma$,
the derivatives of $\vec\Lambda(\vec r)$ with respect to $\vec\tau$
and $\vec\nu^+$ are
\begin{align}
\partial_{\vec\tau}\vec\Lambda(\vec r)&=
 K_{\vec\tau}\vec J_{\rm s}(\vec r)\,,
\qquad\vec r\in\Gamma\,,\\
\partial_{\vec\nu^+}\vec\Lambda(\vec r)&=
\frac{1}{2}\vec J_{\rm s}(\vec r)
+K_{\vec\nu}\vec J_{\rm s}(\vec r)\,,
\qquad\vec r\in\Gamma\,,
\end{align}
where the operators 
\begin{align}
K_{\vec\tau}g(\vec r)&=
\int_\Gamma\vec\tau\cdot\nabla\Phi_k(\vec r,\vec r')
g(\vec r')\,{\rm d}\Gamma'\,,\\
K_{\vec\nu} g(\vec r)&=
\int_\Gamma\vec\nu \cdot\nabla\Phi_k(\vec r,\vec r')
g(\vec r')\,{\rm d}\Gamma'\,,
\label{eq:Knu}
\end{align}
introduced in~\cite{HelsKarl14}, are of the double-layer
type~(\ref{eq:Kdef}). In particular,
\begin{align}
D_{\vec\nu}(\vec r,\vec r')&=-\frac{\left(\nu\cdot(r-r')+\nu_\rho\rho'
\left(1-\cos(\theta-\theta')\right)\right)}
{4\pi|{\vec r}-{\vec r}'|^3}\,,\\
D_{\vec\nu n}(r,r')&=\eta\left[
d(\nu)\mathfrak{R}_n(\chi)-\frac{\nu_\rho}{2\rho}
\left(\mathfrak{R}_n(\chi)+\mathfrak{Q}_{n-\frac{1}{2}}(\chi)\right)\right]\,.
\end{align}

The integral representation of $\Psi(\vec r)$ in terms of $\vec J_{\rm
  s}(\vec r)$ is given by~(\ref{eq:Psidef}) with~(\ref{eq:scud}). In
order to avoid the numerical differentiation associated
with~(\ref{eq:scud}) we follow~\cite{Vicoetal13} and derive a Fredholm
second kind integral equation for $\varrho_{\rm s}(\vec r)$ based on
the observation that
\begin{align}
\lim_{V\ni\vec r\to\vec r^\circ}\vec\nu^\circ
\cdot\vec E(\vec r)&=-\varrho_{\rm s}(\vec r^\circ)\,, \qquad
\vec r^\circ\in\Gamma\,,
\label{Efield1}\\
\vec E(\vec r)&={\rm i}k\vec\Lambda(\vec r)
-\nabla\Psi(\vec r)\,, \qquad\vec r\in\mathbb{R}^3\setminus\Gamma\,,
\label{Efield2}
\end{align}
where $\vec E(\vec r)$ is the electric field divided by the free space
wave impedance. From~(\ref{eq:Psidef}), \eqref{eq:Knu}, and
\eqref{Efield2} we get
\begin{equation}
\lim_{V\ni \vec r\to\vec r^\circ}\vec\nu^\circ\cdot\vec E(\vec r)=
{\rm i}k\vec\Lambda_{\vec\nu}(\vec r^\circ)
-\frac{1}{2}\varrho_{\rm s}(\vec r^\circ)
-K_{\vec\nu}\varrho_{\rm s}(\vec r^\circ)\,, \qquad\vec r^\circ\in\Gamma\,,
\end{equation}
and combine this with~\eqref{Efield1} to obtain an integral equation
which in modal form reads
 \begin{equation}
\left(I-2\sqrt{2\pi}K_{\vec\nu n}\right)\varrho_{{\rm s}n}(r)=
-2{\rm i}k\Lambda_{\nu n}(r)\,, \qquad r\in\gamma\,.
\label{eq:ECCIE}
\end{equation}

From~(\ref{eq:BC}), expressed in terms of $\vec E(\vec r)$ as
\begin{equation}
\lim_{V\ni\vec r\to\vec r^\circ}\vec\nu^\circ\times\vec E(\vec r)=
\vec 0\,,\qquad\vec r^\circ\in\Gamma\,,
\end{equation}
and \eqref{Efield2} we obtain
\begin{align}
\partial_{\vec\tau}\Psi(\vec r)
&={\rm i}k \Lambda_\tau(\vec r)\,,\qquad\vec r\in\Gamma\,,\\
\partial_{\vec\theta}\Psi(\vec r)
&={\rm i}k\rho \Lambda_\theta(\vec r)\,,\qquad\vec r\in\Gamma\,.
\end{align}
The Fourier coefficients of these relations read
\begin{align}
\left(\partial_{\vec\tau}\Psi\right)_n\!(r)
&={\rm i}k\Lambda_{\tau n}(r)\,,\qquad r\in\gamma\,,\\
n\Psi_n(r)
&=\rho k\Lambda_{\theta n}(r)\,,\qquad r\in\gamma\,,
\label{psicoeff}
\end{align}
and are used for the evaluation of
$\left(\partial_{\vec\tau}\Psi\right)_n\!(r)$ and of $\Psi_n(r)$,
$n\ne 0$. 

Explicit expressions in terms of $J_{\tau n}(r)$ and $J_{\theta n}(r)$
for the coefficients of $\vec\Lambda(\vec r)$, with basis as
in~(\ref{eq:Lttn}), are
\begin{align}
\Lambda_{\tau n}(r)&=\sqrt{2\pi}S_{1n}J_{\tau n}(r)
             +{\rm i}\sqrt{2\pi}S_{2n}J_{\theta n}(r)\,,\\
\Lambda_{\theta n}(r)&={\rm i}\sqrt{2\pi}S_{3n}J_{\tau n}(r)
                             +\sqrt{2\pi}S_{4n}J_{\theta n}(r)\,,\\
\Lambda_{\nu n}(r)&=\sqrt{2\pi}S_{5n}J_{\tau n}(r)
            +{\rm i}\sqrt{2\pi}S_{6n}J_{\theta n}(r)\,,
\end{align}
where 
\begin{equation}
S_{\alpha n}g_n(r)=
\int_\gamma S_{\alpha n}(r,r')g_n(r')\rho'\,{\rm d}\gamma'\,, 
\qquad \alpha=1,\ldots,6\,,
\end{equation}
and $S_{\alpha n}(r,r')$ are transformed kernels of operators
$S_\alpha$ of the form~(\ref{eq:Sdef}). The static kernels
$Z_\alpha(\vec r,\vec r')$ are
\begin{align}
Z_1(\vec r,\vec r')&=
 \left(\nu_z\nu_{z}'\cos(\theta-\theta')+\nu_{\rho}\nu_{\rho}'\right)
 \Phi_0(\vec r,\vec r')\,,\\
Z_2(\vec r,\vec r')&=
-{\rm i}\nu_z\sin(\theta-\theta')\Phi_0(\vec r,\vec r')\,,\\
Z_3(\vec r,\vec r')&=
 {\rm i}\nu_z'\sin(\theta-\theta')\Phi_0(\vec r,\vec r')\,,\\
Z_4(\vec r,\vec r')&=
 \cos(\theta-\theta')\Phi_0(\vec r,\vec r')\,,\\
Z_5(\vec r,\vec r')&=
 \left(\nu_\rho\nu_z'\cos(\theta-\theta')-\nu_z\nu_\rho'\right)
 \Phi_0(\vec r,\vec r')\,,\\
Z_6(\vec r,\vec r')&=
-{\rm i}\nu_\rho\sin(\theta-\theta')\Phi_0(\vec r,\vec r')\,,
\end{align}
with closed-form Fourier coefficients
\begin{align}
Z_{1n}(r,r')&=\eta\left[\frac{\nu_z\nu'_z}{2}\left(
\mathfrak{Q}_{n-\frac{3}{2}}(\chi)+\mathfrak{Q}_{n+\frac{1}{2}}(\chi)\right)
+\nu_\rho\nu'_\rho\mathfrak{Q}_{n-\frac{1}{2}}(\chi)\right]\,,\\
Z_{2n}(r,r')&=-\eta\frac{\nu_z}{2}\left(
\mathfrak{Q}_{n-\frac{3}{2}}(\chi)-\mathfrak{Q}_{n+\frac{1}{2}}(\chi)\right)
\,,\\
Z_{3n}(r,r')&=\eta\frac{\nu'_z}{2}\left(
\mathfrak{Q}_{n-\frac{3}{2}}(\chi)-\mathfrak{Q}_{n+\frac{1}{2}}(\chi)\right)
\,,\\
Z_{4n}(r,r')&=\eta\frac{1}{2}\left(
\mathfrak{Q}_{n-\frac{3}{2}}(\chi)+\mathfrak{Q}_{n+\frac{1}{2}}(\chi)\right)
\,,\\
Z_{5n}(r,r')&=\eta\left[\frac{\nu_\rho\nu'_z}{2}\left(
\mathfrak{Q}_{n-\frac{3}{2}}(\chi)+\mathfrak{Q}_{n+\frac{1}{2}}(\chi)\right)
-\nu_z\nu'_\rho\mathfrak{Q}_{n-\frac{1}{2}}(\chi)\right]\,,\\
Z_{6n}(r,r')&=-\eta\frac{\nu_\rho}{2}\left(
\mathfrak{Q}_{n-\frac{3}{2}}(\chi)-\mathfrak{Q}_{n+\frac{1}{2}}(\chi)\right)\,.
\end{align}

For $\Psi_0(r)$ and for $\left(\partial_{\vec\nu^+}\Psi\right)_n\!(r)$
we use
\begin{align}
\Psi_0(r)&=\sqrt{2\pi}S_{\varsigma 0}\varrho_{{\rm s}0}(r)\,, 
\qquad r\in\gamma\,,\\
\left(\partial_{\vec\nu^+}\Psi\right)_n\!(r)&=
\varrho_{{\rm s}n}(r)+{\rm i}k\Lambda_{\nu n}(r)\,,
\qquad r\in\gamma\,,
\end{align}
with $\varrho_{{\rm s}n}(r)$ from~(\ref{eq:ECCIE}).

\section{Code for toroidal harmonics}
\label{app:toro}

The {\sc Matlab} function {\tt toroharm} is a modification of the
standard {\sc Matlab} function {\tt ellipke}. It returns accurate
values of $\mathfrak{Q}_{n-\frac{1}{2}}(\chi)$ for $n=0,1$
\begin{verbatim}
  function [QA,QB]=toroharm(chi,b0)
  a0 = 1;
  s0 = 0;
  i1 = 0.5; 
  w1 = 1;
  while max(w1(:)) > eps
    w1 = i1*(a0-b0).^2;
    s0 = s0+w1;
    a1 = a0+b0;
    b0 = sqrt(a0.*b0);
    a0 = a1/2;
    i1 = 2*i1;  
  end
  mu = sqrt(2./(chi+1));
  QA = pi*mu./a1;
  QB = pi*s0./(mu.*a1);
\end{verbatim}
The output arguments {\tt QA} and {\tt QB} correspond to
$\mathfrak{Q}_{-\frac{1}{2}}(\chi)$ and
$\mathfrak{Q}_{\frac{1}{2}}(\chi)$. The input argument {\tt chi}
corresponds to $\chi$ and the input argument {\tt b0} should be chosen
as
\begin{equation}
{\tt b0}=\sqrt{\frac{\chi-1}{\chi+1}}\,.
\label{eq:b0}
\end{equation}
The reason for providing {\tt b0} as an extra input argument is that
this quantity may be available to higher relative precision than what
comes from a direct evaluation via $\chi$ and~(\ref{eq:b0}).
Compare~(\ref{eq:chidef}) when $|r-r'|$ is small.

\begin{small}

\end{small}

\end{document}